%% file: g346_ms.tex
\documentclass[iop]{emulateapj} 
\submitted{Accepted to ApJ} 
\usepackage[colorlinks=true,linkcolor=blue,citecolor=blue]{hyperref}
\usepackage{color}
\usepackage{graphicx}
\usepackage[dvipsnames]{xcolor}

\begin{document}

\title{AN \emph{XMM-NEWTON} STUDY OF THE MIXED-MORPHOLOGY SUPERNOVA REMNANT G346.6--0.2}

\author{Katie Auchettl\altaffilmark{1,2,$\dagger$}, C-Y. Ng\altaffilmark{3,*}, B.T.T. Wong\altaffilmark{3}, Laura Lopez\altaffilmark{1,4}, Patrick Slane\altaffilmark{5}}

\altaffiltext{1}{Center for Cosmology and AstroParticle Physics (CCAPP), The Ohio State University, 191 W. Woodruff Ave., Columbus, OH 43210, USA}
\altaffiltext{2}{Department of Physics, The Ohio State University, 191 W. Woodruff Ave., Columbus, OH 43210, USA}
\altaffiltext{3}{Department of Physics, The University of Hong Kong, Pokfulam Road, Hong Kong, China}
\altaffiltext{4}{Department of Astronomy, The Ohio State University, 191 W. Woodruff Ave., Columbus, OH 43210, USA}
\altaffiltext{5}{Harvard-Smithsonian Center for Astrophysics, 60 Garden Street, Cambridge, MA 02138, USA}
\altaffiltext{$\dagger$}{auchettl.1@osu.edu}
\altaffiltext{*}{ncy@bohr.physics.hku.hk}

\begin{abstract}

We present an X-ray imaging and spectroscopic study of the
molecular cloud interacting mixed-morphology (MM) supernova remnant (SNR)
G346.6--0.2 using \emph{XMM-Newton}. The X-ray spectrum of the remnant is well described by a recombining plasma that most likely arises from adiabatic cooling, and has sub-solar abundances of Mg, Si, and S. Our fits also suggest the presence of either an additional power-law component with a photon index of $\sim$2, or an additional thermal component with a temperature of $\sim$2.0 keV. We investigate the possible origin of this component and suggest that it could arise from either the Galactic ridge X-ray emission, an unidentified pulsar wind nebula or X-ray synchrotron emission from high-energy particles accelerated at the shock. However, deeper, high resolution observations of this object are needed to shed light on the presence and origin of this feature. Based on its morphology, its Galactic latitude, the density of the surrounding environment and its association with a dense molecular cloud, G346.6--0.2 most likely arises from a massive progenitor that underwent core-collapse.

\end{abstract}


\section{Introduction}
Supernova remnants (SNRs) are structures that result from the explosive
end of massive stars. The energy from the supernova explosion is
partially converted into kinetic energy and is dissipated in collisionless
shocks that heat the stellar ejecta and swept-up interstellar medium (ISM) to
X-ray emitting temperatures. Apart from sweeping up and heating material, the
shock-front of SNRs are sites where relativistic particles
can be efficiently accelerated to energies up to $10^{15}$\,eV
(i.e., the ``knee'' of the Cosmic-ray spectrum) \citep{2014ChPhC..38i0001O}. Non-thermal
X-ray emission arising from shock-accelerated particles has been detected in a handful of SNRs ($\sim 14$ out of 294 known Galactic SNRs). This emission is found to originate predominantly from the shell of the remnant (e.g., SN1006:
\citealt{1995Natur.378..255K}; RX J1713.7$-$3946:
\citealt{1997PASJ...49L...7K,1999ApJ...525..357S}; and Vela Jr.:
\citealt{1998Natur.396..141A,2001ApJ...548..814S}), or in thin filaments at
the edges of young SNRs (e.g., Tycho: \citealt{2002ApJ...581.1101H,
2005ApJ...634..376W}; and Kepler: \citealt{2004A&A...414..545C}), and detection of this emission provides
direct evidence for electrons being accelerated to TeV energies. While the
non-thermal X-rays in the above remnants are confined to narrow regions close
to the shock front, this does not seem to be the case for somewhat older and
physically much larger SNRs like RCW~86, Vela Jr., and RX J1713.7-3946,
in which the emission regions are broader and located behind the shock
\citep{2000PASJ...52.1157B, 2001ApJ...548..814S, 2004A&A...427..199C}. A
number of these synchrotron X-ray emitting SNRs such as RCW~86 and Tycho also
emit noticeable thermal X-ray emission from both ejecta and shocked
circumstellar material that surround the non-thermal X-ray filaments. 

All non-thermal emitting SNRs detected so far are classified as shell-like, and none of these remnants are known to be interacting with nearby molecular clouds through the detection of 1720\,MHz OH maser(s)
\citep{1999AJ....117.1387C}\footnote{Even though there are a number of methods (see \citealt{2015SSRv..188..187S} for more details) used to infer the presence of SNR/MC interaction, the detection of an OH maser is a ``smoking gun'' signal, since they can only be formed in conditions related to a shock/molecular cloud interaction. A handful of these non-thermal SNRs such as RX J1713.7$-$3946 \citep{1999ApJ...525..357S, 2001ApJ...562L.167B} show evidence of shock interaction in the form of other molecular line features, however none of these sources so far show evidence of an OH maser.}. On the other hand, a large fraction of X-ray emitting SNRs are
known to be interacting with molecular clouds and are classified as mixed
morphology (MM) SNRs \citep{1998ApJ...503L.167R}.  Unlike shell-type SNRs
whose X-ray emission traces a shell, MM SNRs have a centrally peaked X-ray
morphology that arises from a collisionally heated plasma located in the interior of the radio shell, while often showing enhanced elemental abundances, and isothermal temperatures (e.g., \citealt{2006ApJ...647..350L}). 

The morphology and
X-ray properties of these remnants are unexpected if one assumes
standard SNR evolution models (e.g., \citealt{1977ARA&A..15..175C,
1999ApJS..120..299T, 2000ApJS..128..403T}), and the evolutionary process which
lead to these characteristics are not well understood. There are two main
models which attempt to explain the properties of MM SNRs. The first one is the
thermal conduction model in which heat and material are transported to the centre of the remnant via the Coulomb collisions between electrons and ions inside the hot plasma,
resulting in the centrally-filled emission and isothermal temperatures \citep{1992ApJ...401..206C, 1999ApJ...524..179C}. The second one invokes
the evaporation of clumps of material that are sufficiently small and dense to not be destroyed by or disrupted by the the shock itself
\citep{1991ApJ...373..543W}. Some MM SNRs also
show evidence of overionisation which results from the rapid cooling of electrons, and manifests itself in the form of recombination edges
\cite[e.g.,][]{2005ApJ...631..935K, 2009ApJ...706L..71O, 2013ApJ...777..145L}. This rapid cooling can occur either by adiabatic expansion
\citep[e.g.,][]{1989MNRAS.236..885I}, thermal conduction \citep{2002ApJ...572..897K} or the interaction with dense cavity walls or molecular clouds \citep{2005ApJ...630..892D}.

Here we present an analysis of a molecular cloud interacting MM SNR
G346.6--0.2 which shows evidence of a hard X-ray tail component. 

G346.6--0.2 was first discovered and classified as a shell-type SNR in the
480\,MHz and 5\,GHz radio bands \citep{1975AuJPA.......75C} using the
the Molonglo Observatory Synthesis Telescope (MOST) and the Parkes 64-m radio
telescope. The shell-like morphology of the remnant was confirmed
using the Very Large Array (VLA) and MOST at 1465\,MHz and at 843\,MHz, respectively
\citep{1993AJ....105.2251D,1996A&AS..118..329W}. It has an angular
diameter of $8\farcm2\pm0\farcm5$ and a radio-continuum spectral index of $-0.6\pm0.1$
\citep{2001ApJ...559..963G}. The eastern and north-western edges of the
remnant show evidence of interaction with the surrounding environment, such as the flattening of the radio contours in these regions \citep{1993AJ....105.2251D}.
A number of 1720\,MHz OH masers were detected along the southern rim of the
radio shell at a velocity of $-76.0$\,km\,s$^{-1}$
\citep{1998AJ....116.1323K}, indicating that the SNR is interacting with
surrounding molecular clouds. \citet{1998AJ....116.1323K}
calculated the kinematic distance towards the masers (and thus the SNR) using
the Galactic rotation curve. They determined a distance of
5.5\,kpc and 11\,kpc, respectively, with a tangent point distance of 8.3
kpc in this direction. Using the $\Sigma$-$D$ relationship
\citep{1985ApJ...295L..13H} a distance of 9\,kpc to G346.6--0.2 was suggested
\citep{1993AJ....105.2251D}. In this paper, we use 8.3 kpc as
the distance to the remnant, similar to that of previous studies.

In X-rays, G346.6--0.2 was first detected in the \emph{ASCA} Galactic plane
survey \citep{2008PASJ...60.1143Y}. The \emph{ASCA} GIS image shows centrally-filled
diffuse X-ray emission, indicating that this is an MM SNR. Although
photon statistics were limited, \citet{2008PASJ...60.1143Y} determined that
the X-ray spectrum could be modelled using a thermal plasma (MEKAL model) with
a temperature of $\sim$1.6 keV or a power-law model with a photon index of
$\sim$3.7. They also derived a column density of $N_{\rm H} \sim
(2-2.6)\times10^{22}$ cm$^{-2}$. \citet{2014AJ....147...55P} re-analysed the
\emph{ASCA} data using different models (PHABS$\times$POWERLAW,
PHABS$\times$APEC, PHABS$\times$NEI and combinations of these). They found
that the X-ray spectrum is best described by an absorbed non-equilibrium
ionisation (NEI) model with a column density of $N_{\rm H} =
2.1^{+0.4}_{-0.7}\times10^{22}$\,cm$^{-2}$, temperature of
$2.8^{+1.1}_{-0.5}$\,keV, and an ionisation timescale of
$7^{+6}_{-4}\times10^9$\,cm$^{-3}$\,s. Based on \textit{Suzaku} observations,
it was claimed that the X-ray spectrum can be fitted by an absorbed hot ($kT =
1.22\pm0.04$\,keV) NEI model with sub-solar abundances of Mg, Si, S, and Fe,
plus a power-law with a photon index of $0.6\pm0.3$
\citep{2011MNRAS.415..301S}. However, a later reanalysis of the same data,
after properly accounting for the strong X-ray emission from the Galactic
Ridge, indicates that the X-ray emission is best described by an absorbed
recombining plasma with a temperature of $0.30^{+0.03}_{-0.01}$\,keV,
sub-solar abundances of Mg, Si, S, and Fe and a column density of $N_{\rm H} =
(2.3\pm0.1)\times10^{22}$\,cm$^{-2}$ \citep{2013PASJ...65....6Y}.

The \textit{Spitzer} IRAC survey of SNRs in the inner Galaxy
\citep{2006AJ....131.1479R} detected diffuse infrared emission arising from
the southern rim of the radio shell of G346.6--0.2. This IR emission is
coincident with the OH masers detected towards the south and the IRAC colours
derived in this region suggest molecular cloud interaction. There is also
fainter infrared emission towards the northern edge of the remnant. The
detection of spectral lines associated with shocked H$_2$ emission from
G346.6--0.2 using \textit{Spitzer} IRS observations
\citep{2009ApJ...694.1266H} indicates that the remnant is interacting with a
high density environment such as a molecular cloud. However, 
no $\gamma$-ray emission was found, using approximately 3.5\,years
of PASS 7 \textit{Fermi}-LAT data \citep{2012AIPC.1505..265E}.

In this paper, we present an X-ray observation of SNR G346.6--0.2 using
\textit{XMM-Newton}. In Section 2, we describe the \textit{XMM} data reduction, spatial and spectral analysis and our point source analysis. In Section 3 and 4 we infer the properties of G346.6--0.2, and discuss the origin and nature of its thermal and non-thermal X-ray emission, while in Section 5 we discuss the possible origin of the non-thermal component. In Section 6 we discuss the nature of the point sources we detected and search for a potential neutron star candidate, while in Section 7 we summarise our results. 

\begin{figure*}[!th]
	\begin{center}
		\includegraphics[width=0.47\textwidth]{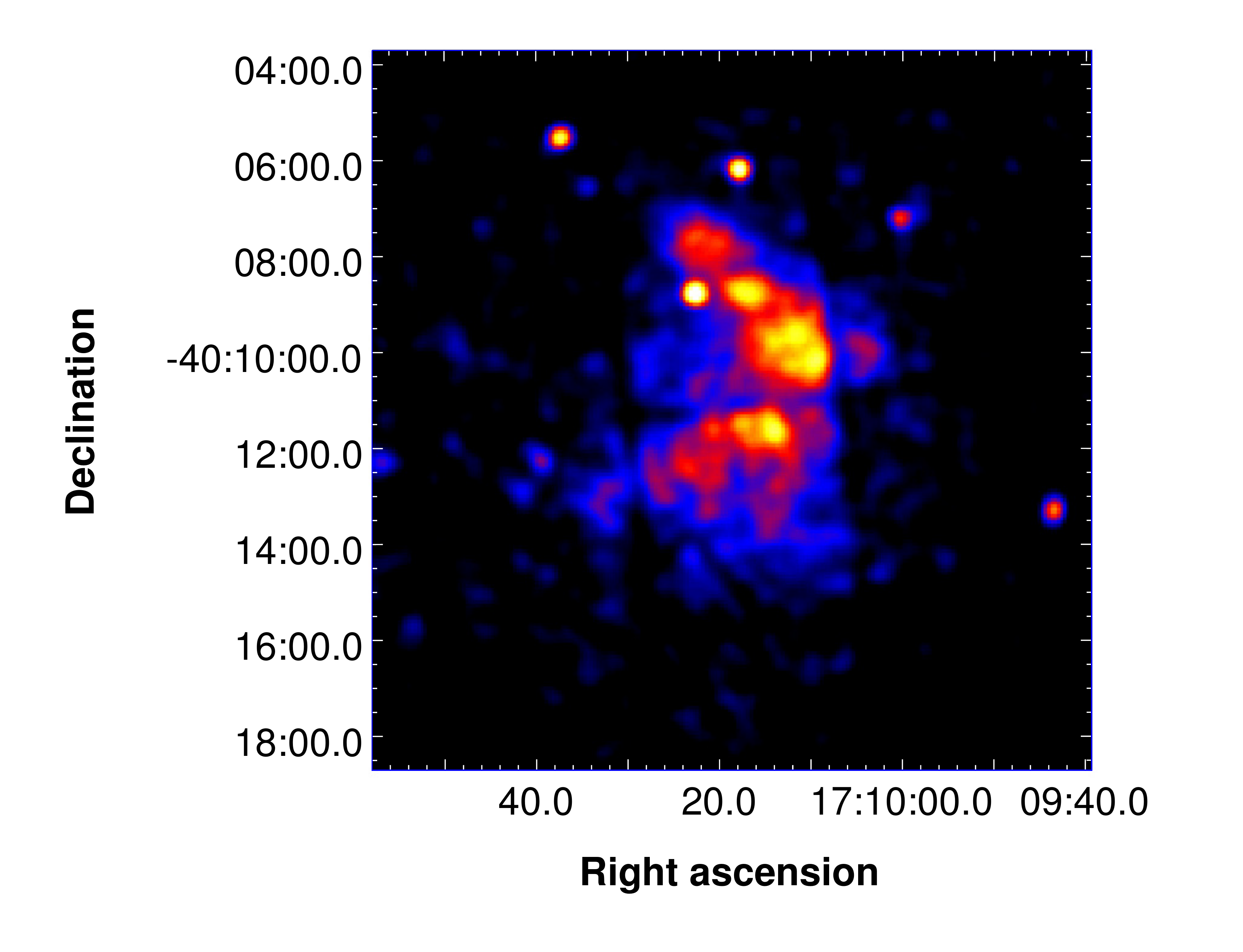}
		\includegraphics[width=0.46\textwidth]{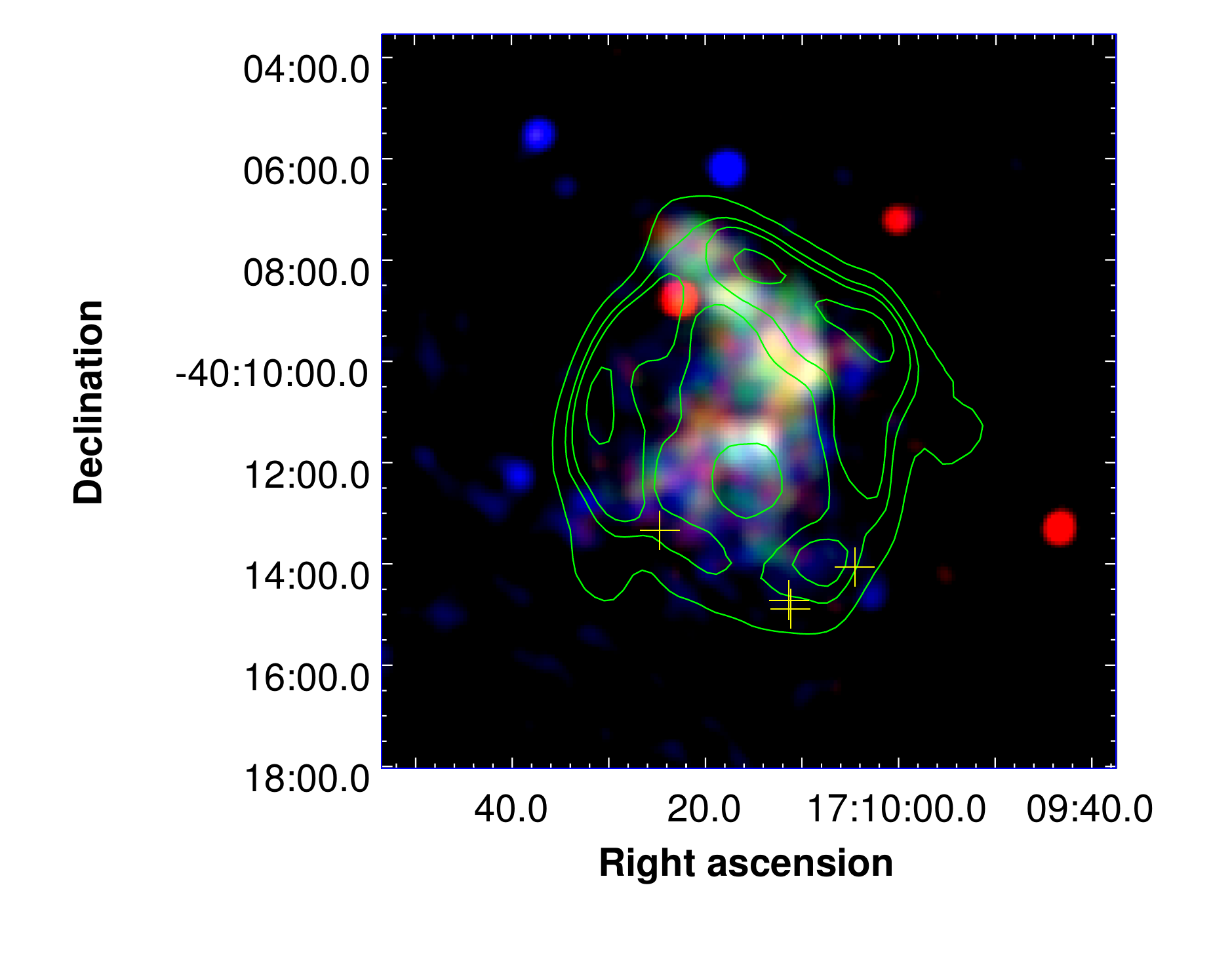}
		\caption{\textit{Left:} Exposure corrected, 0.5--7.0\,keV MOS+PN mosaic
image of the G346.6--0.2. The image is smoothed with a Gaussian of
width 20\arcsec\ and the colour scale is linear.  The apparent sub-luminous nature of central X-ray emission of the remnant is a result of the chip gap of the PN detector rather that a property of the remnant (see Figure \ref{pointsrc} where this effect is more pronounced).  \textit{Right:} RGB image of G346.6--0.2 using both the
MOS and the PN cameras. Red corresponds to 0.5--1.5\,keV, green to 1.5--2\,keV
and blue to 2--7\,keV. The 843\,MHz radio contours from MOST are shown in
green \citep{1996A&AS..118..329W}. The image is smoothed with a Gaussian of
width 20\arcsec. The yellow crosses indicate the location of the 1720\,MHz 
OH masers detected with the VLA \citep{1998AJ....116.1323K}. \label{xrayimage}}
\end{center}
\end{figure*}

\section{\textit{XMM-Newton} Observations, Analysis, and Results}
SNR G346.6--0.2 was observed with both the MOS and PN detectors onboard the
\textit{XMM-Newton} Observatory on 2011 March 11 for a total of 30.1 ks
(ObsID: \dataset[ADS/Sa.XMM#obs/0654140101]{0654140101}). The telescope was pointed at
$\rm (\alpha,\delta)=(17^h09^m59.8^s, -40^\circ12'56.4'')$ and the
MOS and PN detectors were operated in the full frame mode with the thick filter, with the SNR fully enclosed by the field of view of the detectors.
We performed the data reduction and analysis using the
\textit{XMM-Newton} science system (SAS) version
14.0.0\footnote{\url{https://www.cosmos.esa.int/web/xmm-newton/documentation/}} with
CALDB 4.6.7\footnote{\url{https://www.cosmos.esa.int/web/xmm-newton/calibration}}.

Before completing our imaging and spectral analysis, we first checked for
periods of high background and/or proton flares by generating a count rate
histogram using events with energy between 10--12\,keV for the observation.
We find that our observation is only slightly affected by high
background or flares, giving effective exposures of 29.4\,ks, 29.3\,ks and
24.6\,ks for MOS1, MOS2 and PN, respectively. As suggested in the SAS analysis
threads\footnote{\url{https://www.cosmos.esa.int/web/xmm-newton/sas-threads}}
and \textit{XMM-Newton} Users
Handbook\footnote{\url{https://xmm-tools.cosmos.esa.int/external/xmm_user_support/documentation/sas_usg/USG/}},
we reduced the data following the standard screening of events, with single to
quadruple events (PATTERN $\le$ 12) chosen for the MOS detectors, while for
the PN detector only single and double events (PATTERN $\le$ 4) were selected.
We also used the standard canned screening set of FLAGS for
both the MOS (\#XMMEA\_EM) and PN (\#XMMEA\_EP) detector. 

As G346.6--0.2 is located along the Galactic plane, both Galactic Ridge X-ray Emission (GRXE) and the Cosmic X-ray Background (CXB) can contribute non-negligibly to the observed emission. To correct for this we must take vignetting effects into account. Thus, we process all event files using the task \texttt{evigweight}, which weights each event by an energy dependent factor that is equivalent to the ratio of the effective area at the centre of the observation and the effective area at the position of interest\footnote{See \url{https://xmm-tools.cosmos.esa.int/external/sas/current/doc/evigweight} for more details}. All analysis products and results presented below are extracted from these cleaned, filtered and vignetting corrected event files.

\subsection{Imaging Analysis}
We used the SAS task \texttt{emosaic} to combine the MOS and PN observations to
produce a single exposure-corrected intensity image of the entire SNR. The
resulting image in the 0.5--7\,keV energy band is shown in Figure
\ref{xrayimage} (left). To determine any possible spectral variations of the
remnant and the nature of point sources in the field, we also generated an RGB
image using events from both the MOS and PN detectors (Figure
\ref{xrayimage} right).

These images reveal that the X-ray emission is relatively clumpy
in nature. The bulk of the X-ray emission found towards the north of the
remnant produces a significant amount of soft X-rays, while the emission seen
towards the southern region overlapping the positions of the OH masers is
harder in nature. The X-ray emission from the SNR shows an arc-like morphology and it is brightest towards the west. The emission extends in both the northeast and southeast, with the northern extension seeming to follow the slight protrusion of the radio contours, as seen in the RGB image (Figure \ref{xrayimage}, right). The X-ray emission of this remnant is
surrounded by faint diffuse emission and is fully enclosed by the MOST radio
contours \citep{1996A&AS..118..329W}, covering a region with an approximate size of $7\farcm3\times8\farcm2$. 

There are a few bright point sources
immediately surrounding the SNR (see Figure 1). One of them coincides with the radio
shell in the north of the remnant, and another is found in the southwest near the OH masers. All other point sources are found outside the radio remnant shell. The point sources surrounding the remnant emit strongly in either soft or hard
X-rays only. In particular, the point source within the northern part of the radio shell is dominated by soft X-rays. A more detailed analysis and discussion
of the point sources in the field of view are found in Section
\ref{pointsources}. 

\begin{figure}[!t]
	\includegraphics[width=\columnwidth]{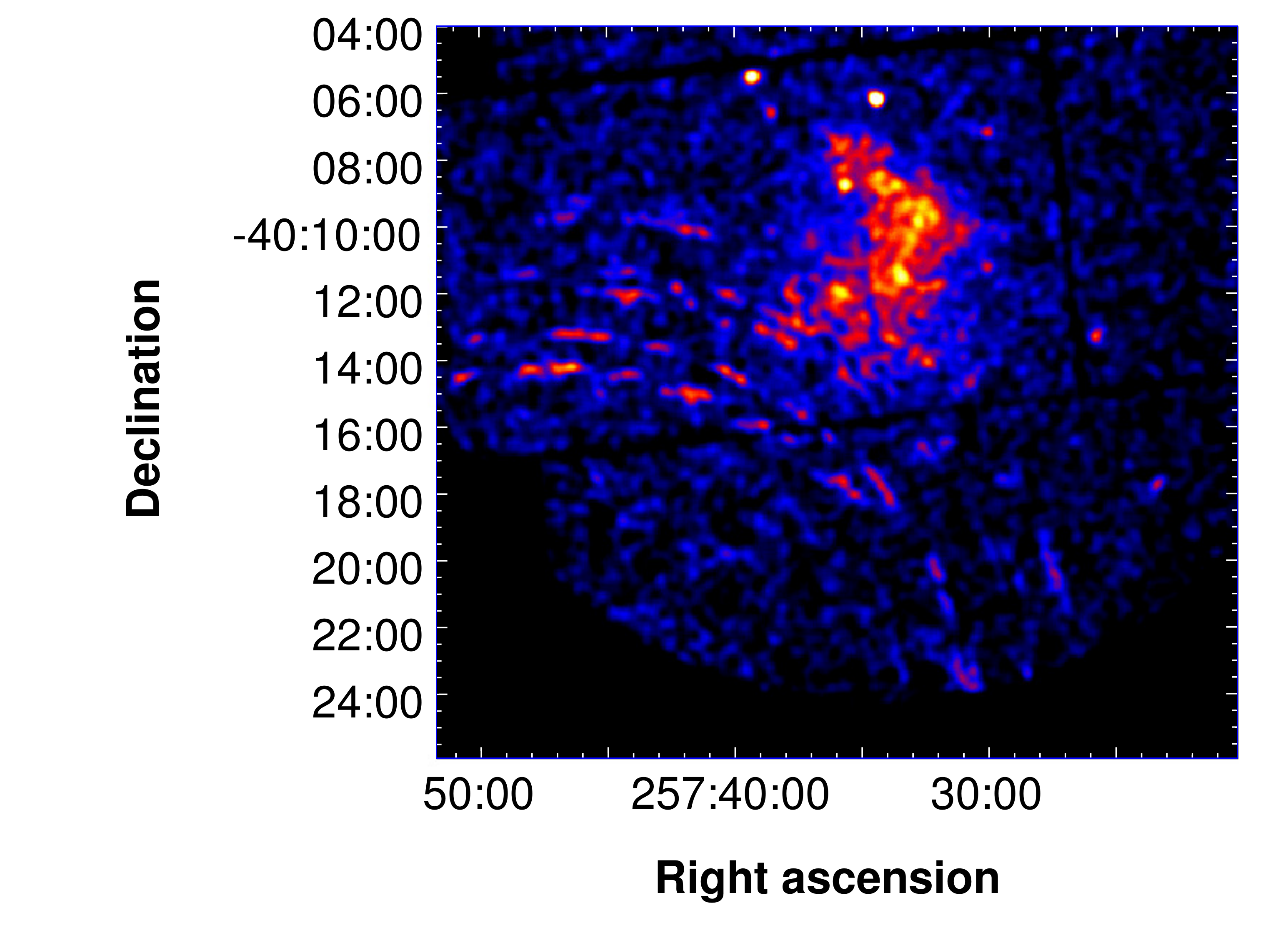}
	\begin{center}
		\caption{MOS2 image of G346.6--0.2 showing the singly-reflected X-rays from a nearby source in the form of large arcs towards the south-east of the remnant. This artefact is seen in all detectors and most prominently in MOS2.\label{reflectedxrays}}
	\end{center}
\end{figure}

We also note that there is an arc-like towards the southeast of the remnant, which could arise from either a bright nearby source or from flaring (see Figure \ref{reflectedxrays}). As we filtered out periods of high-background and/or proton flares before analysing the data products of this observation, it is more likely that these fringes arise from a nearby source. We searched the \textit{ROSAT} All-Sky Survey
Bright Source Catalogue \citep{1999A&A...349..389V} and Third
\emph{XMM-Newton} Serendipitous Source Catalogue \citep{2015arXiv150407051R} and find the low mass X-ray binary 4U $1708-40$ (1RXS J$171224.8-405034$) located approximately $0.8^{\circ}$ from G346.6--0.2, making it likely to be the source of this emission.

\begin{figure*}[!t]
	\begin{center}
			\includegraphics[width=0.45\textwidth]{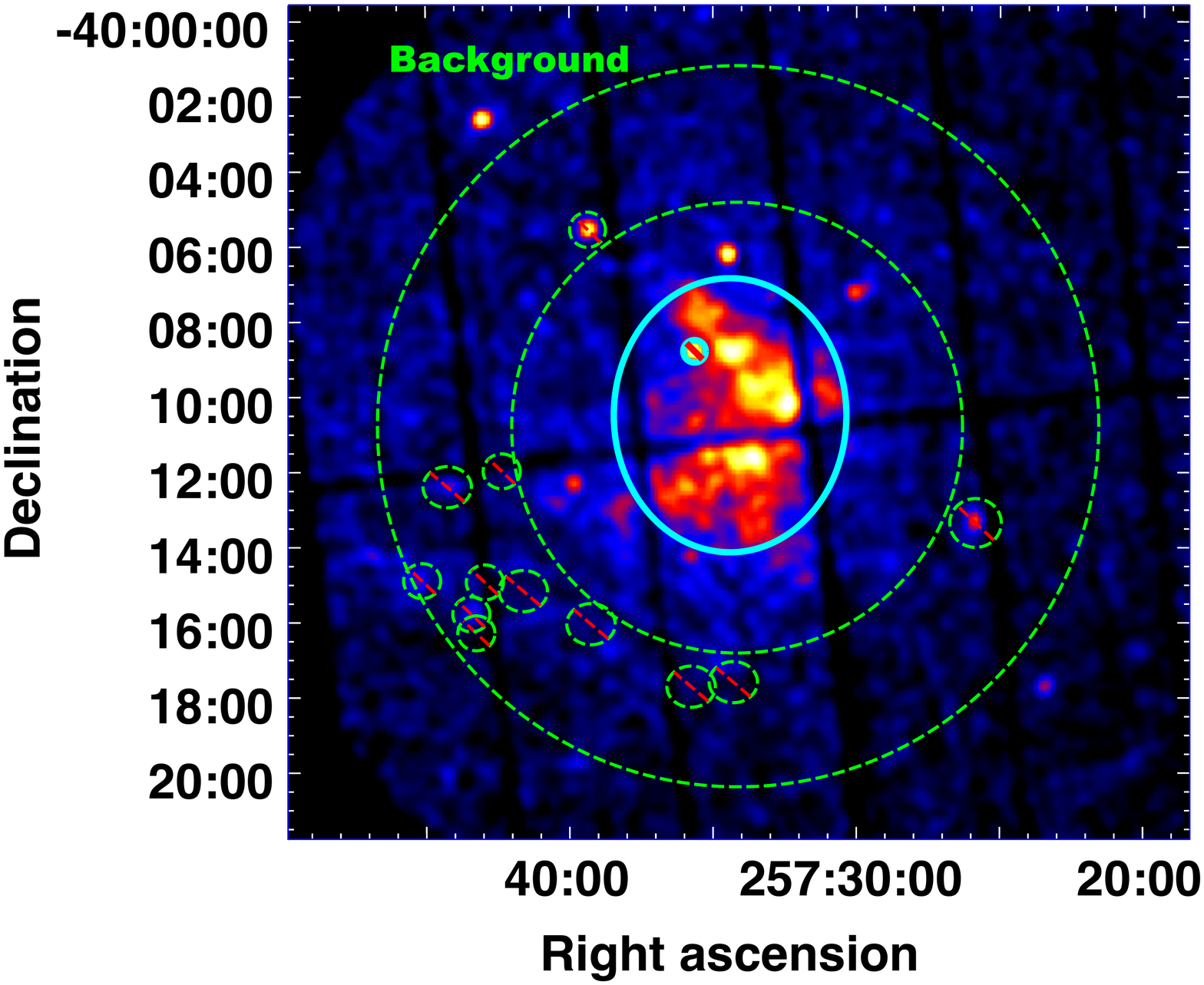}
		\includegraphics[width=0.50\textwidth]{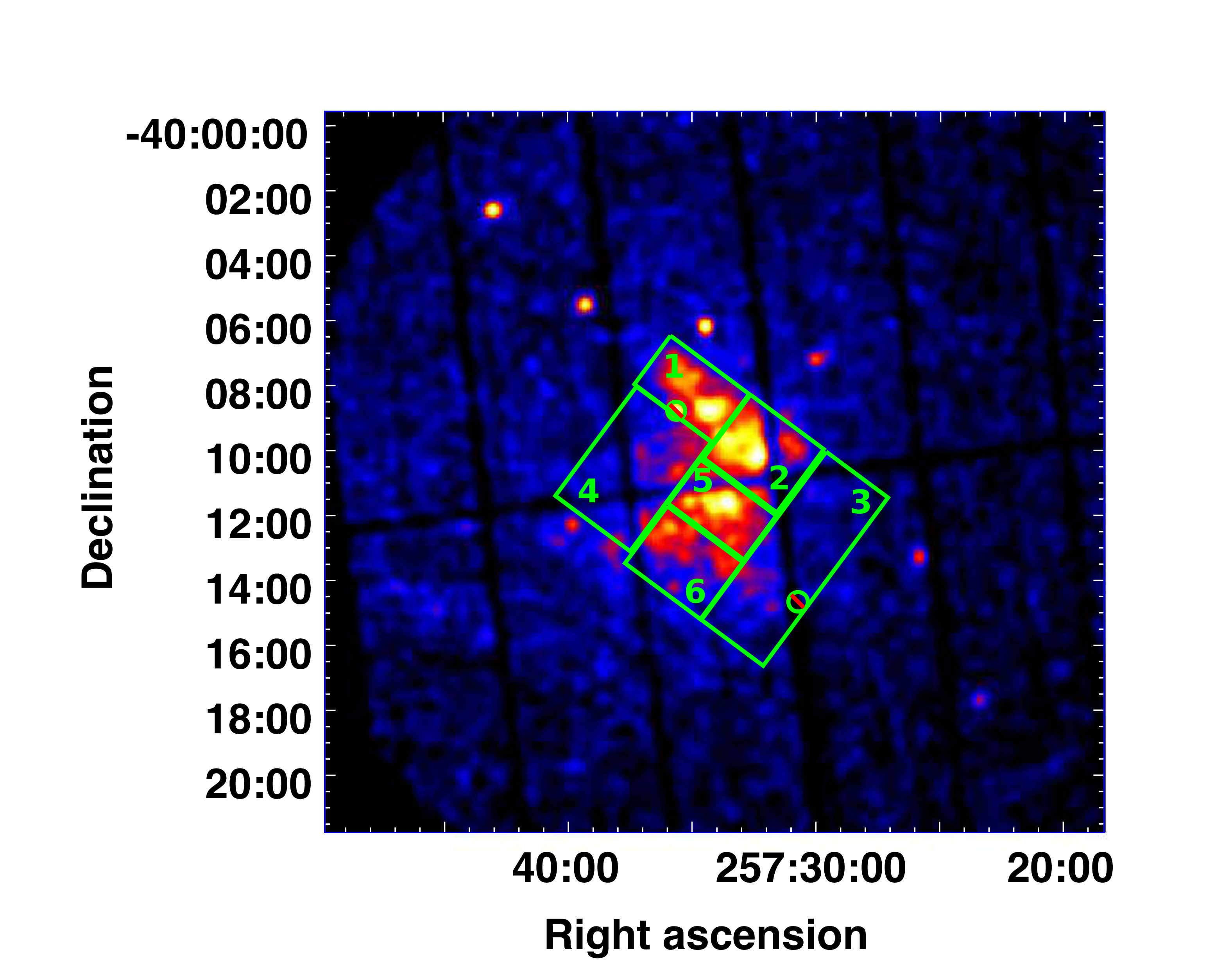}
		\caption{\textit{Right:} Plotted as the solid cyan and dash green regions respectively are the global and background regions used in our spectral analysis overlaid on the 0.5--7.0\,keV MOS+PN mosaic image.  \textit{Left:} The individual source regions used in our analysis. Bright point sources are excluded in the analysis. Here we have adjusted the contrast of the image such that only the brightest emission from the remnant is observed. 
			\label{regions}}
	\end{center}
\end{figure*}

\subsection{Spectral Analysis}\label{spec}

To determine the spectral properties of G346.6--0.2, we extract spectra from
six regions shown in Figure \ref{regions} right using the SAS task
\texttt{evselect} and the cleaned, vignetting corrected event files from all three EPIC cameras. In
addition, we extracted a global spectrum from an elliptical region centred on
$\rm (\alpha,\delta)=(17^{h}10^{m}17.6^{s}, -40^{\circ}10'29.3'')$ with
semi-minor and semi-major radii of 3\arcmin\ and 3.4\arcmin, respectively, which
encloses the bulk of the SNR X-ray emission. Point sources overlapping
these regions were excluded. For each region, we extracted
spectral response and effective area files using the tasks \texttt{arfgen} and
\texttt{rmfgen}.  To account for the background, we selected an annulus region that directly surrounds the emission of the remnant (see Figure \ref{regions} left). We excluded both point source and the arc-shaped fringes from singly-reflected X-rays seen in all observations from our background region.

The spectral fitting was performed using the X-ray analysis software XSPEC
version 12.9.0c, over an energy range of 0.7--7\,keV. We also used AtomDB
3.0.2\footnote{\url{http://www.atomdb.org/}} \citep{2001ApJ...556L..91S,
2012ApJ...756..128F}. Each spectrum was grouped with a minimum of 20 counts
per energy bin, and fitted using $\chi^2$ statistics. To investigate the
emission of the remnant, we used a non-equilibrium ionisation (NEI)
collisional plasma model, VRNEI\footnote{\url{https://heasarc.gsfc.nasa.gov/xanadu/xspec/manual/node210.html}}, which is characterised by final ($kT$) and
initial electron temperature ($kT_{\rm init}$), elemental abundances and a
single ionisation timescale ($\tau = n_{e}t$). This model
allows one to simulate the thermal emission from either a plasma that is
ionising up to or is in ionisation equilibrium (i.e.\ $kT_{\rm init} < kT$), or a
recombining plasma that was in collisional equilibrium with $kT_{\rm
init}$ and then suddenly cooled to its final temperature $kT$ (i.e.\ $kT_{\rm
init} > kT$ ). We also used an absorbed APEC model \citep{2001ApJ...556L..91S}, which allows one to  model a plasma which is in collisional ionisation equilibrium (CIE).

\begin{figure*}[!ht]
	\begin{center}
		\includegraphics[height=0.49\textwidth,angle=-90]{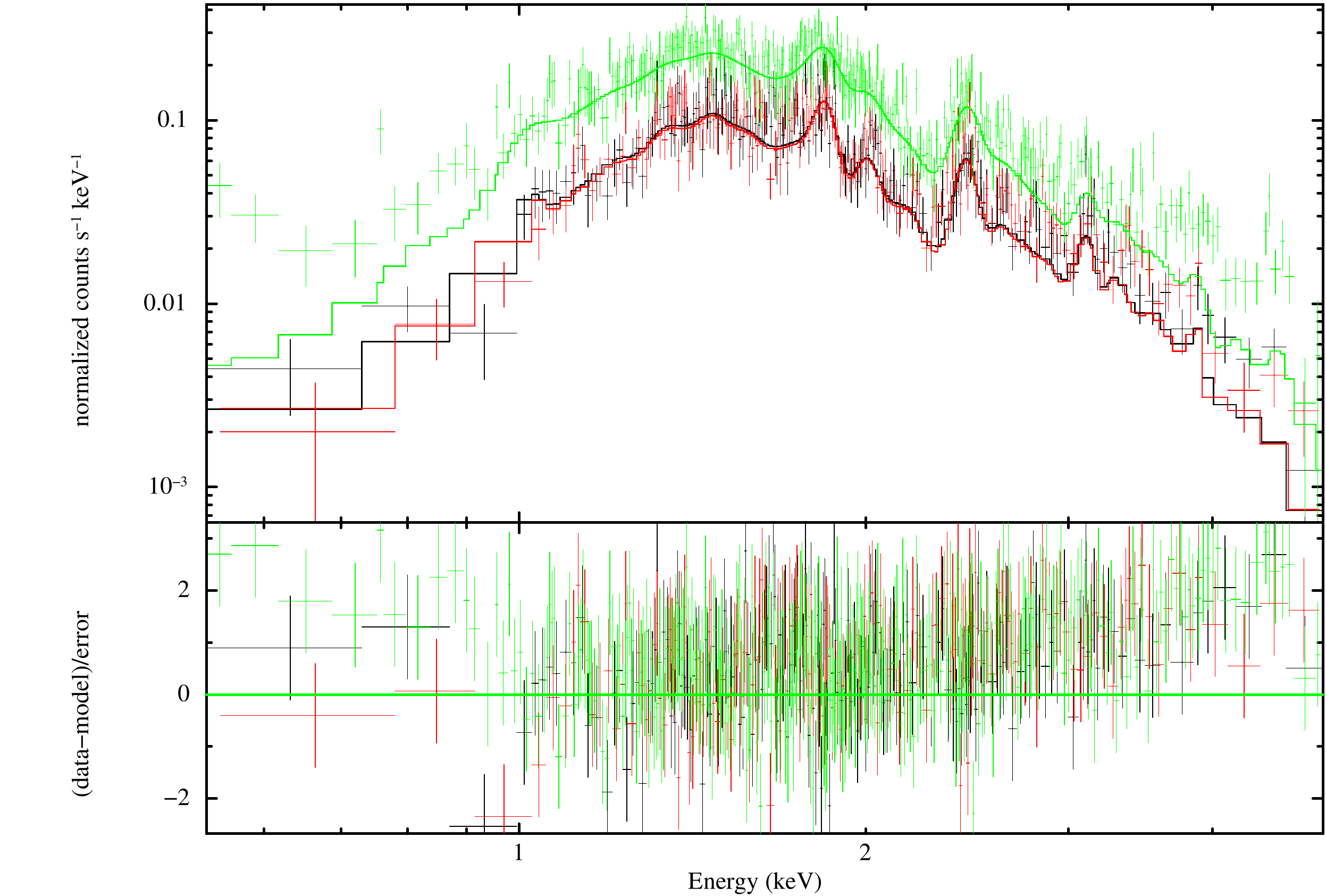}
		\includegraphics[height=0.49\textwidth,angle=-90]{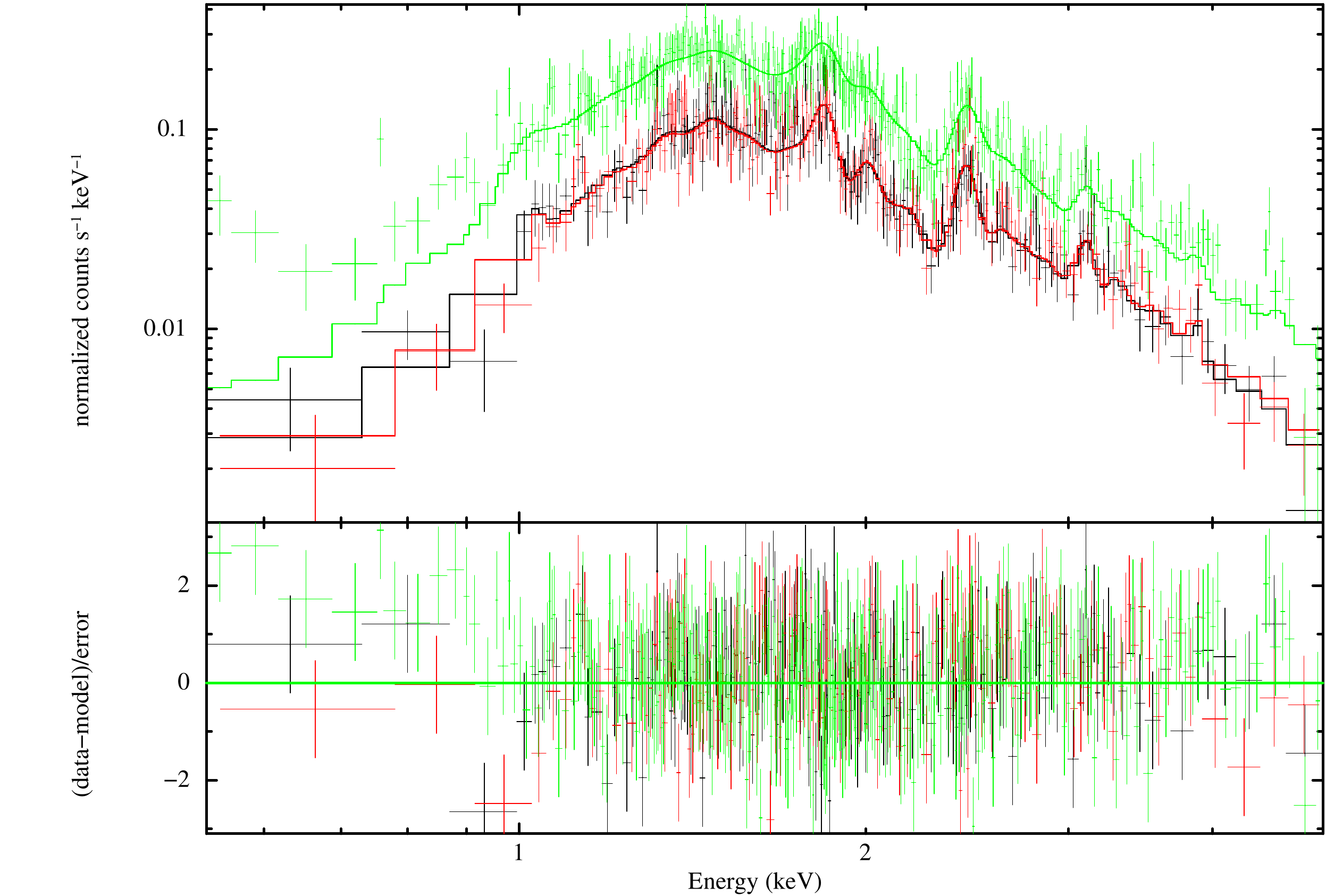}
		\caption{\textit{Left:} Best-fit absorbed VRNEI model for the MOS1 (black), MOS2 (red), and PN (green) spectra, with a  temperature of $0.34\pm0.01$\,keV, $kT_{\rm init}$ fixed at 6\,keV and under-abundant	Mg, Si, S, and Fe. \textit{Right:} Best-fit absorbed VRNEI+power-law model with a temperature of $0.20^{+0.03}_{+0.02}$ keV,  $kT_{\rm init}=$ 6\,keV, under-abundant Mg, Si,  and Si, as well as a power-law index of $\Gamma = 2.3^{+0.5}_{-0.8}$. The left panel shows that 	the single absorbed VRNEI model significantly underestimates the flux above
			$\sim4$\,keV. \label{xrayspectrum}}
	\end{center}
\end{figure*}

The foreground absorption column density $N_{\rm H}$ was modelled using TBABS
\citep{2000ApJ...542..914W}. Due to the presence of noticable emission lines from Mg,
Si, and S, the abundances of these elements were also let free in the VRNEI model fit. All other elemental abundances were fixed at the solar
values \citep{2000ApJ...542..914W}. We find that all regions favoured $kT_{\rm init}>kT$, implying that the X-ray
emission from G346.6--0.2 arises from a recombining plasma, similar to what
was suggested previously \citep{2013PASJ...65....6Y}. For completeness, we also attempted to fit the emission of the remnant using a TBABS$\times$VRNEI model with $kT_{\rm init}<kT$, or an TBABS$\times$APEC model. However, we find that both these models produced a worse fit (reduced $\chi^2\sim 1.3$ for both models), and they suggest extremely high temperatures for the SNR (i.e., $kT_{e}\sim$1.7$-$2.6 keV).  In Table~\ref{fits}, we list the best-fit
parameters and their 90\% confidence level uncertainties, while in
Figure~\ref{xrayspectrum} we plot the global X-ray spectrum of G346.6--0.2. 

We attempt to fit the
value of $kT_{\rm init}$, but found that it is poorly constrained for all individual
regions except for the global spectrum, for which we obtained $kT_{\rm
init}=6^{+4}_{-1}$\,keV. Therefore, we fixed $kT_{\rm init}$ at the
global value when fitting the spectra from individual regions. 

Overall, a TBABS$\times$VRNEI model with under-abundant Mg, Si, S, and Fe, a
plasma temperature of $kT\sim0.30$\,keV and $kT_{\rm init}$ fixed at
6\,keV, was able to fit the global spectrum reasonably well with a reduced $\chi^2 = 1.16$. This is consistent with that derived by \citet{2013PASJ...65....6Y} using \emph{Suzaku} data.  However, this model fails to fit the high energy
spectrum above $\sim$3\,keV, as shown by the residuals in Figure
\ref{xrayspectrum} (left panel). The \emph{Suzaku} X-ray spectrum and model fits presented in Table 1 of  \citet{2013PASJ...65....6Y} also hints at an additional spectral component \footnote{The best-fit recombining NEIJ model presented by \citet{2013PASJ...65....6Y} has a reduced $\chi^{2}$ between 1.20$-$1.40 depending on the background they used.,
but the poor statistics above $\sim$4\,keV precludes a detailed study of this component in their paper.} To account
for this excess seen in the hard X-ray band (Figure \ref{xrayspectrum} left panel), we attempted to add a recombining plasma (RNEI), or an NEI model component, however we found that these models did not improve the fit when $\tau<10^{12} \rm cm^{-3} s$. However, we find that these models significantly improve the fit when $\tau>10^{13} \rm cm^{-3} s$, mimicking that of an APEC model which we discuss further below.

It is possible that this hard X-ray tail could arise from either excess thermal emission from the Galactic Ridge or from a powerlaw component that arises from a non-thermal population of the electrons accelerated by the supernova shock or from an unseen pulsar wind nebula (PWN). To test these possibilities, we added either an additional APEC model \citep{2001ApJ...556L..91S} or a powerlaw model in which we let both the temperature and the photon index be free respectively. 

We find that adding a powerlaw component significantly improves the fit and gives a reduced
$\chi^2$ of 0.99, and an $F$-test indicates a null hypothesis probability of
$1.0\times10^{-15}$ when compared with our best fit TBABS$\times$VRNEI model. The same is also true for individual spectra except region 2, all have an $F$-test
null hypothesis probability of $<10^{-5}$. The powerlaw component for the global spectrum has a photon index of $\Gamma = 2.0^{+0.7}_{-0.9}$, while the individual regions have $\Gamma$
between 1.0 and 2.5 (see Table \ref{fits}). The power-law index implied by our fits is similar to that obtained by \citet{2008PASJ...60.1143Y} whose power-law
model fit of \emph{ASCA} observations of the remnant required a power-law index
$>1.7$. These values are much larger than $\Gamma\sim0.5$ obtained with
\emph{Suzaku} using a WABS*(VNEI+POWERLAW) model \citep{2011MNRAS.415..301S}, however these authors did not correct for vignetting. In Table \ref{fits} we have listed the results of our fits using either a TBABS$\times$VRNEI or TBABS$\times$(VRNEI+POWERLAW) model.

Similar to our powerlaw model, we find that an additional APEC model also improves the fit to our global and individual spectra (reduced $\chi^2$ of 0.96 for the global spectrum). Here the best fit APEC temperature derived from the global spectrum is $1.9_{-0.3}^{+0.6}$ keV, while those for the individual regions are similar, albeit with larger uncertainties (see Table \ref{fitsapec}). This temperature is much higher than that expected from a SNR, or from the 0.79 keV excess emission seen directly surrounding G346.6--0.2 \citep{2013PASJ...65....6Y}, while it is lower than that derived by \citet{2014AJ....147...55P} using a single APEC model. 

The derived temperature value is higher than that estimated by various studies of the Galactic Ridge Emission using X-rays \citep[e.g.,][]{2012ApJ...753..129Y, 2014PASJ...66..125Y, 2013PASJ...65...19U, 2016ApJ...833..268N}, however, within uncertainties it is consistent with that of the low temperature component of the Galactic Ridge Emission derive by e.g., \citet{2012ApJ...753..129Y} and \citet{2013PASJ...65...19U}. We find that the properties of the VRNEI models in these fits are comparable to those in the VRNEI+POWERLAW fits (see Table \ref{fits}).

We find that both the global and individual regions have an ionisation
timescale of $\sim4.8\times10^{11}$\,cm$^{-3}$\,s, indicating that the plasma
is far from ionisation equilibrium. This is similar to the values derived by
\citet{2013PASJ...65....6Y} and \citet{2011MNRAS.415..301S}, but is
significantly different from the \emph{ASCA} and \textit{Suzaku} results
\citep{2008PASJ...60.1143Y, 2014AJ....147...55P}. The average temperature of the recombining plasma is $0.26$\,keV, which is slightly lower than the $0.30^{+0.03}_{-0.01}$ keV reported by \citet{2013PASJ...65....6Y} and lower than the temperature of 0.30 keV derived using only a TBABS$\times$VRNEI model.

As \citet{2008PASJ...60.1143Y},
\citet{2011MNRAS.415..301S}, and \citet{2014AJ....147...55P} used a CIE model or a non-equilibrium ionisation model without overionisation to describe the recombining plasma, our results differ from their studies. In addition, \citet{2008PASJ...60.1143Y} and \citet{2011MNRAS.415..301S} did not properly account for either
the non-X-ray background of \textit{Suzaku} or vignetting effects, which can lead to the plasma
temperature being much higher in their model fits to compensate for the excess
flux above $\sim 5$\,keV \citep{2013PASJ...65....6Y}.  Another difference is
that we use updated elemental abundances \citep{2000ApJ...542..914W} and
atomic cross-sections (ATOMDB 3.0.2) in the
analysis, compared to much older versions (e.g., \citealt{1989GeCoA..53..197A} abundances or ATOMDB 2.0) used in previous studies.

The global and individual spectra, except for region 3, all suggest
under-abundance of Mg. In addition, the global and regions 1, 2 and 4 spectra
require under-abundance of Si. Under-abundance of S is required for the global
and regions 1, 2, 4 and 5 spectra. This is similar to the \emph{Suzaku}
results \citep{2011MNRAS.415..301S,2013PASJ...65....6Y}.
However, we do not find evidence for under-abundance of Fe or over abundance
of Ca as claimed by the above two studies. Note that the \emph{ASCA} data were
unable to verify these due to limited photon statistics
\citep{2008PASJ...60.1143Y,2014AJ....147...55P}.

We derived a hydrogen column density of $N_{\rm H}=(2.0-3.3)\times
10^{22}$\,cm\,$^{-2}$. It is highest in regions 1 and 4 in the eastern side of
the remnant, and lowest on the western side (region 3). The $N_{\rm H}$ values are
comparable to those previously reported \citep{2008PASJ...60.1143Y,
2011MNRAS.415..301S,2013PASJ...65....6Y,2014AJ....147...55P}.

Finally, to check the possibility that the additional component could arise from the singly reflected X-rays seen in Figure \ref{reflectedxrays}, we estimate the flux contribution
of this artefact using WebPIMMS\footnote{\url{https://heasarc.gsfc.nasa.gov/cgi-bin/Tools/w3pimms/w3pimms.pl}}. Assuming the $N_{\rm H}$ and $\Gamma$ derived from
our global X-ray spectrum, this corresponds to an absorbed flux of
$\sim1\times10^{-14}$\,erg cm$^{-2}$ s$^{-1}$ over the 0.5--7.0 keV energy band.
This is only $\sim$2\% of the absorbed flux of the power-law component in
both the global spectrum and in all regions which require an additional
power-law.

\subsection{Point Source Analysis}\label{pointsources}
We identified X-ray point sources in the field of view by
running the task \texttt{edetect\_chain} on the PN data in
five standard energy bands (0.2--0.5\,keV, 0.5--1\,keV, 1--2\,keV,
2--4.5\,keV, and 4.5--10\,keV). We chose a likelihood threshold value of
30$\sigma$ to mitigate background contamination. A total of 25 bright point
sources were detected in the field. Their positions and count rates in the
soft (0.2--2\,keV) and hard (2--10\,keV) bands are listed in
Tables~\ref{counterparts} and \ref{ptsrctable}. We calculated the  hardness ratio (HR) using
($R_{2-10}-R_{0.2-2}$)/($R_{2-10}+R_{0.2-2}$), where $R$ are the count rates
in the soft and hard bands from the PN detector. We use only the events from PN detector for this
analysis due to its high sensitivity and large effective area. Also, a few
sources fall off the MOS detectors, or are located on CCD6 of MOS1
which is no longer operational. 

For sources which we were able to extract
sufficient counts and thus extract an X-ray spectrum, we grouped the spectra with a minimum of 10 counts per bin, and modelled each spectrum using an absorbed power-law. We compared
the positions of the X-ray sources with UNSO-B1 \citep{2003AJ....125..984M}
and 2MASS \citep{2006AJ....131.1163S} catalogues to identify any optical or
infrared counterparts, respectively. We considered only optical and infra-red
sources within 3$\sigma$ uncertainties of the X-ray positions, and identified
14 optical counterparts. No optical counterparts are found within 3$\sigma$ uncertainties for the other 11 sources. The results of our point source analysis are listed in Tables~\ref{counterparts} and~\ref{ptsrctable}.

\begin{figure*}[!t]
	\begin{center}
		\includegraphics[width=0.33\textwidth]{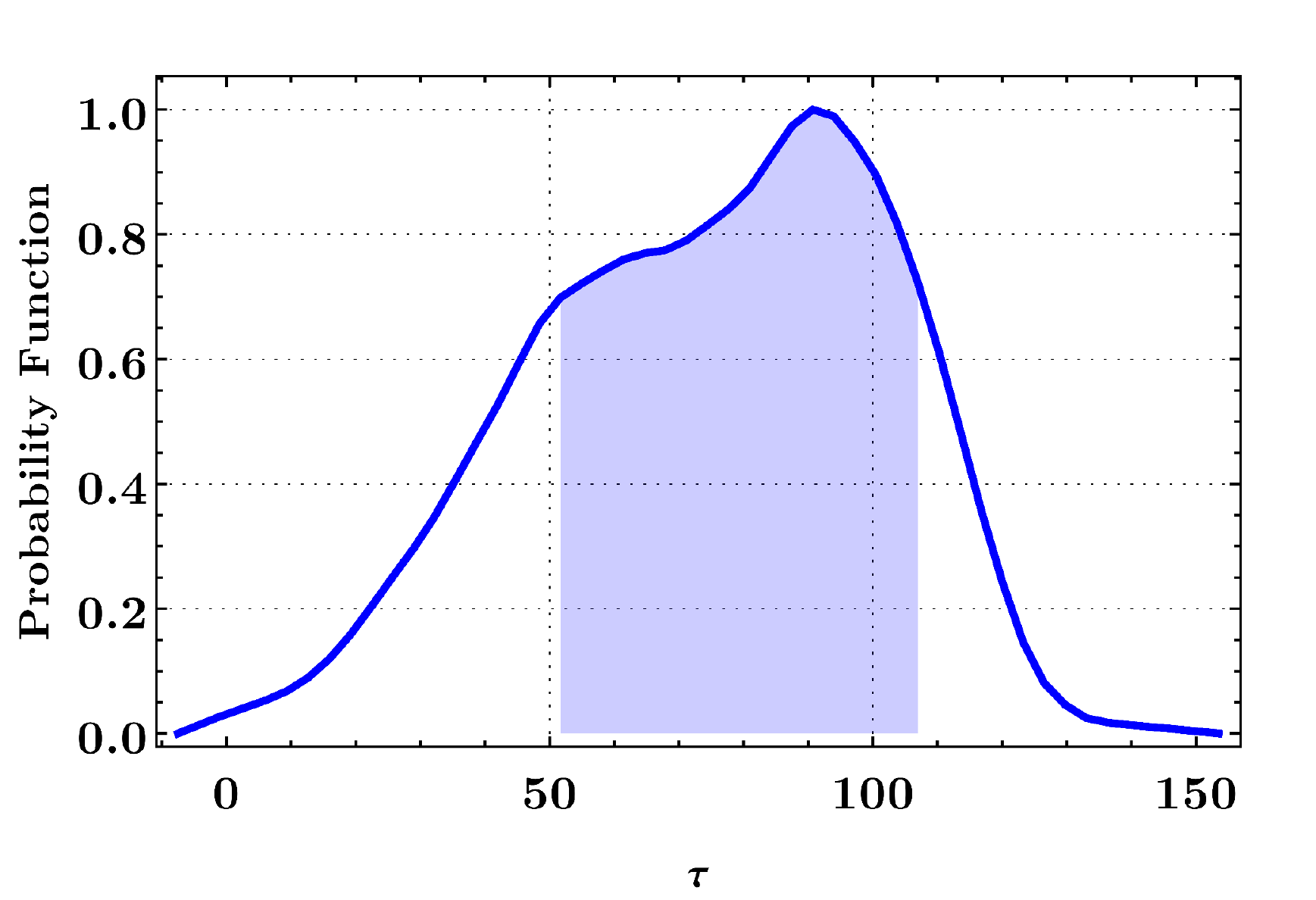}
		\includegraphics[width=0.33\textwidth]{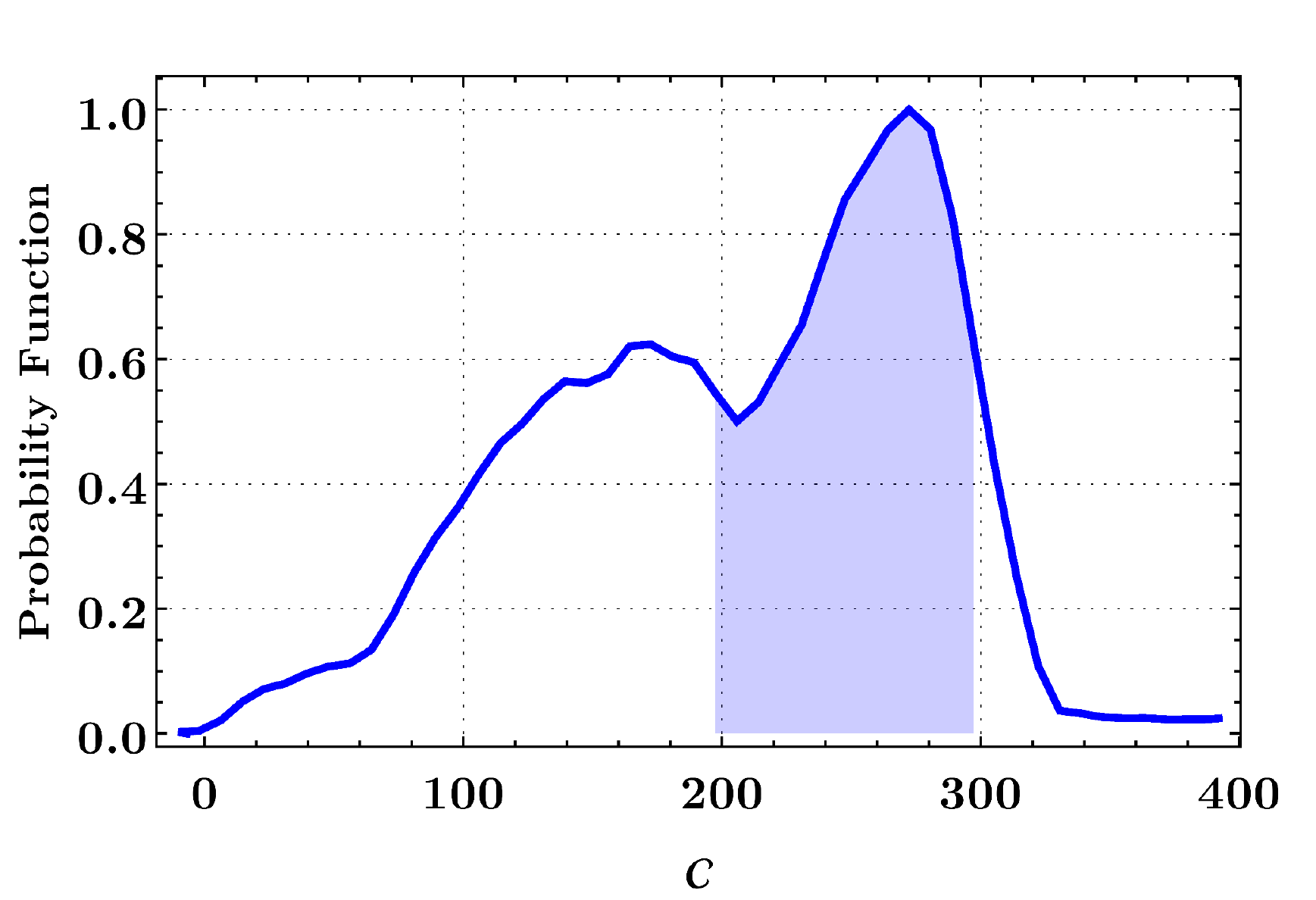}
		\includegraphics[width=0.33\textwidth]{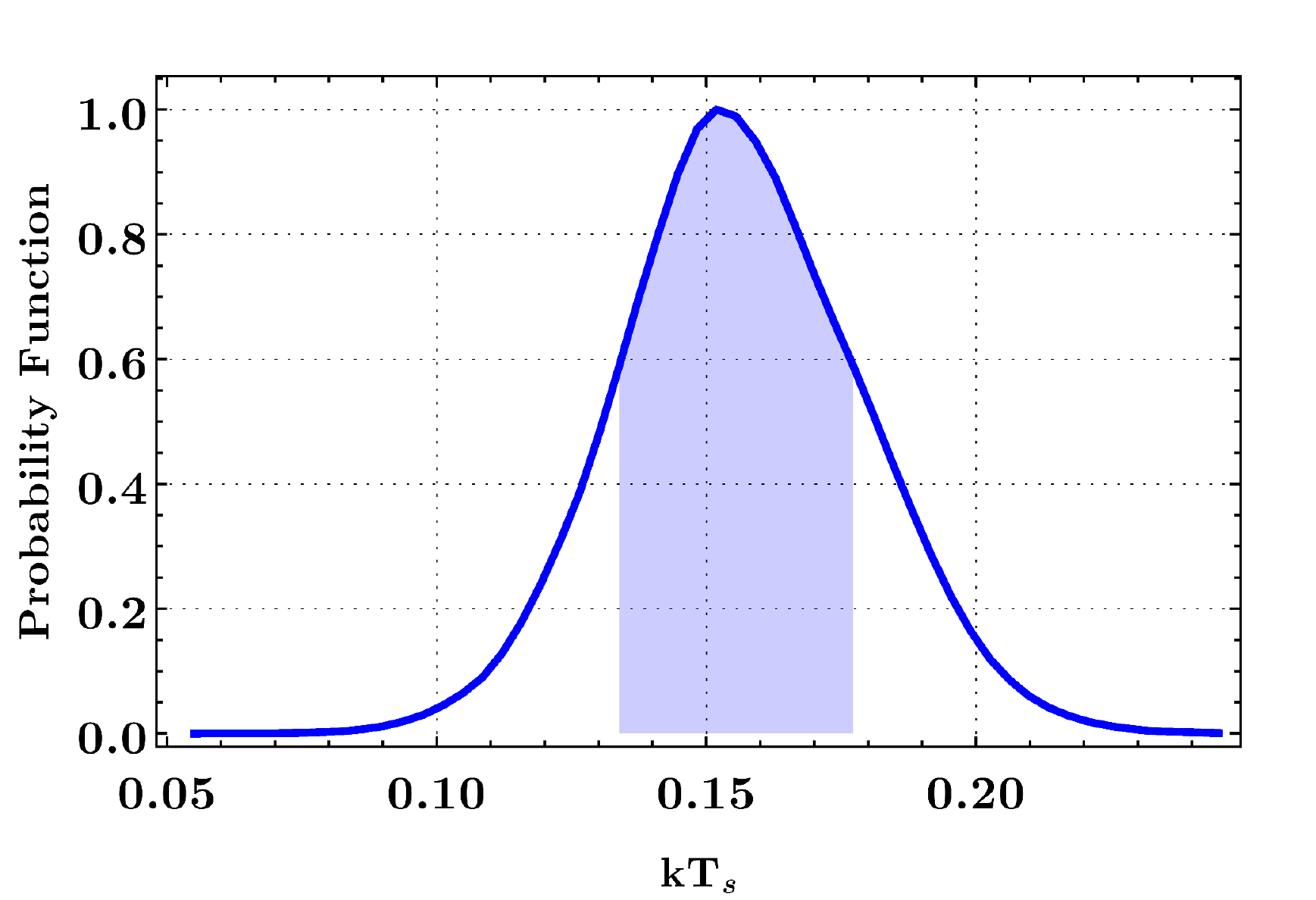}
		\caption{Probability distributions of $kT_{s}$, $\mathcal{C}$ and $\tau$ from the \citet{1991ApJ...373..543W} model, where the shaded regions show the 1$\sigma$ confidence intervals for each parameter. The best fit model parameters for G346.6$-$0.2 are ($\tau$,  $\mathcal{C}$, $kT_{s}$) = ($94^{+23}_{-36}, 260^{+63}_{-96}, 0.16\pm0.02$ keV).
			\label{posteriors}}
	\end{center}
\end{figure*}

\section{Possible origin of the centre-filled X-ray emission}\label{ages}

The detection of OH masers coincident with G346.6$-$0.2 implies that this remnant is expanding into a dense
and non-uniform environment. The characteristics of MM SNRs are thought to arise from the interaction between the SNR with this dense
molecular material. However, their morphological and plasma properties are difficult to explain using the standard Sedov SNR evolution model
\citep[see e.g.,][]{1991ApJ...373..567L, 2000ApJ...545..922S}. Two popular models\footnote{There are other models which attempt to explain the properties of MM SNRs. \citet{2001A&A...371..267P} suggest that the central X-ray emission of MM SNRs arises from projection affects, while \citet{2005ApJ...630..892D} suggested that their properties arise from the collision of the SNR blast wave with the dense cavity walls or ring-like structures that are produced by massive progenitors as they lose mass. However a detailed comparision between models is beyond the scope of this paper.} used to explain the properties of MM SNRs are thermal conduction \citep[e.g.,][]{1992ApJ...401..206C,
1999ApJ...524..179C} and the evaporation of dense clumps of material inside
the remnant \citep[e.g.,][]{1991ApJ...373..543W}.  

In the thermal conduction model \citep{1992ApJ...401..206C, 1999ApJ...524..179C}, the  material immediately behind the shock front begins to cool after the passage of the supernova blast wave. If thermal conduction is
the main source of cooling, this results in the transport of heat and material
to the centre of the remnant, increasing its central density and smoothing the
temperature gradient behind the shock. At the shock-front itself, the relatively high density
of the swept-up ISM absorbs X-ray emission from the shell, while only the central emission is observable. This model is able to reproduce the centre bright X-ray emission of a sample of SNRs such as W44 \citep[e.g.,][]{1999ApJ...524..179C}, however it is unable to fully explain the temperature and brightness distribution of other MM SNRs.

In the \citet{1991ApJ...373..543W} model, the properties of MM SNRs arises from the evaporation of dense clumps of material. The SNR is assumed to be evolving in a medium that contains many cold cloudlets that are sufficiently small and dense that they
do not affect the passage of the shock and are neither destroyed nor swept up. Once the shock has passed, the cloudlets
are then embedded in the hot post-shock plasma and evaporate via thermal
conduction. This fills the SNR interior with a relatively dense gas that emits
X-rays. 

Even though neither of these
two models are able to reproduce completely the complex emission, morphology and dynamical properties of
all MM SNRs, they provide us with a good handle on their properties
\citep[e.g.,][]{2002ApJ...564..284S, 2013ApJ...777..148G}. As G346.4$-$0.2 is expanding in a highly inhomogeneous medium rather than an environment with a strong density gradient, we adopt the \citet{1991ApJ...373..543W} model to investigate the properties of this remnant.

\subsection{Evaporation of dense clumps}

When the mass of the swept-up ISM is greater than the ejecta mass, the SNR enters the Sedov-Taylor phase \citep{1950RSPSA.201..159T, 1959sdmm.book.....S}. At this point, the ejecta have been shocked and can be approximated assuming that it is adiabatically expanding in a uniform medium. \citet{1991ApJ...373..543W} generalised the Sedov model and its corresponding similarity solution such that it could describe the evolution of SNRs in a dense, non-uniform environment, which takes into account the evaporation of dense clumps of material. \citet{1991ApJ...373..543W} introduce two additional parameters to the Sedov model: $\mathcal{C}$, the ratio of the ISM cloud mass to that of the intercloud medium, and $\tau$, the ratio of the cloud evaporation timescale to the SNR age. For appropriate values of $\mathcal{C}$ and  $\tau$, one is able to reproduce the bright X-ray morphologies of MM SNRs, while also being able to recover the standard \citet{1959sdmm.book.....S} model when $\mathcal{C},\tau\ll1$ or $\tau \gg\mathcal{C}$.

To characterise the properties of G346.6$-$0.2, we sample a wide range of values for the shock temperature  ($kT_{s}$), $\mathcal{C}$ and $\tau$ to determine the \citet{1991ApJ...373..543W} model which best reproduces the properties of the remnant. For each set of parameters we derived the model surface brightness and X-ray temperature profiles assuming the X-ray emissivities of a plasma with an X-ray temperature ($kT_{X}$), ionisation timescale and abundances similar to what we obtained from our spectral analysis of the global X-ray spectrum of the remnant. The X-ray emissivity was calculated assuming the observed column density and using the spectral responses files of the MOS1 detector. For each set of parameters, we calculate the average model X-ray temperature from the corresponding temperature profiles, and compare this result to the X-ray temperature derived from modelling the global X-ray spectrum of the remnant.

We derived the surface brightness profile using the MOS1 observation, as the SNR overlaps with a large number of chip gaps in the PN data. In addition, we used an annulus centred at $\rm (\alpha,\delta)=(17^{h}10^{m}17^{s}, -40^{\circ}10'59'')$ which fully encloses the X-ray emission of the remnant. We also exclude all bright points sources that we found using our point source analysis discussed in Section \ref{pointsources}.

To estimate the values of $kT_{s}$, $\mathcal{C}$ and $\tau$ favoured by the global X-ray temperature and the surface brightness profile of G346.4$-$0.2, we randomly sampled the parameter space of each variable. To determine the best-fit values of these parameters we derived a likelihood function for each variable. A likelihood function, $\mathcal{L}$, is the probability of obtaining the observed data $d_{i}$, given the value of the parameter $p_{n}$ and is derived using
\begin{equation}
\mathcal{L} =  \prod\frac{1}{\sqrt{2\pi}\sigma_{i}}\exp \left(-\frac{\chi_{i}^{2}}{2}\right).
\end{equation}
Here $\chi_{i}$ is the chi-squared fit of our surface brightness model and our model temperature as derived when one compares these values to the X-ray surface brightness profile and global temperature of G346.6--0.2. This is derived using 
$\chi_{i}^{2}=(d_{i}-m_{i})/\sigma_{i}$, where $m_{i}$ is the model prediction assuming parameters $p_{n}$ for data point $d_{i}$, and $\sigma_{i}$ is the uncertainty in $d_{i}$. Assuming that all parameters $p_{i}$ can produce a reasonable fit, we integrate  $\mathcal{L}$ over the full range of parameters, producing a probability function for each parameter $p_{n}$. The 68\% (1$\sigma$) confidence interval for each of the parameters $p_{n}$ are derived using the minimal parameter region that encloses 68\% of the integrated area under the distribution.

\begin{figure*}[!t]
 	\begin{center}
 		
 		\includegraphics[width=0.46\textwidth]{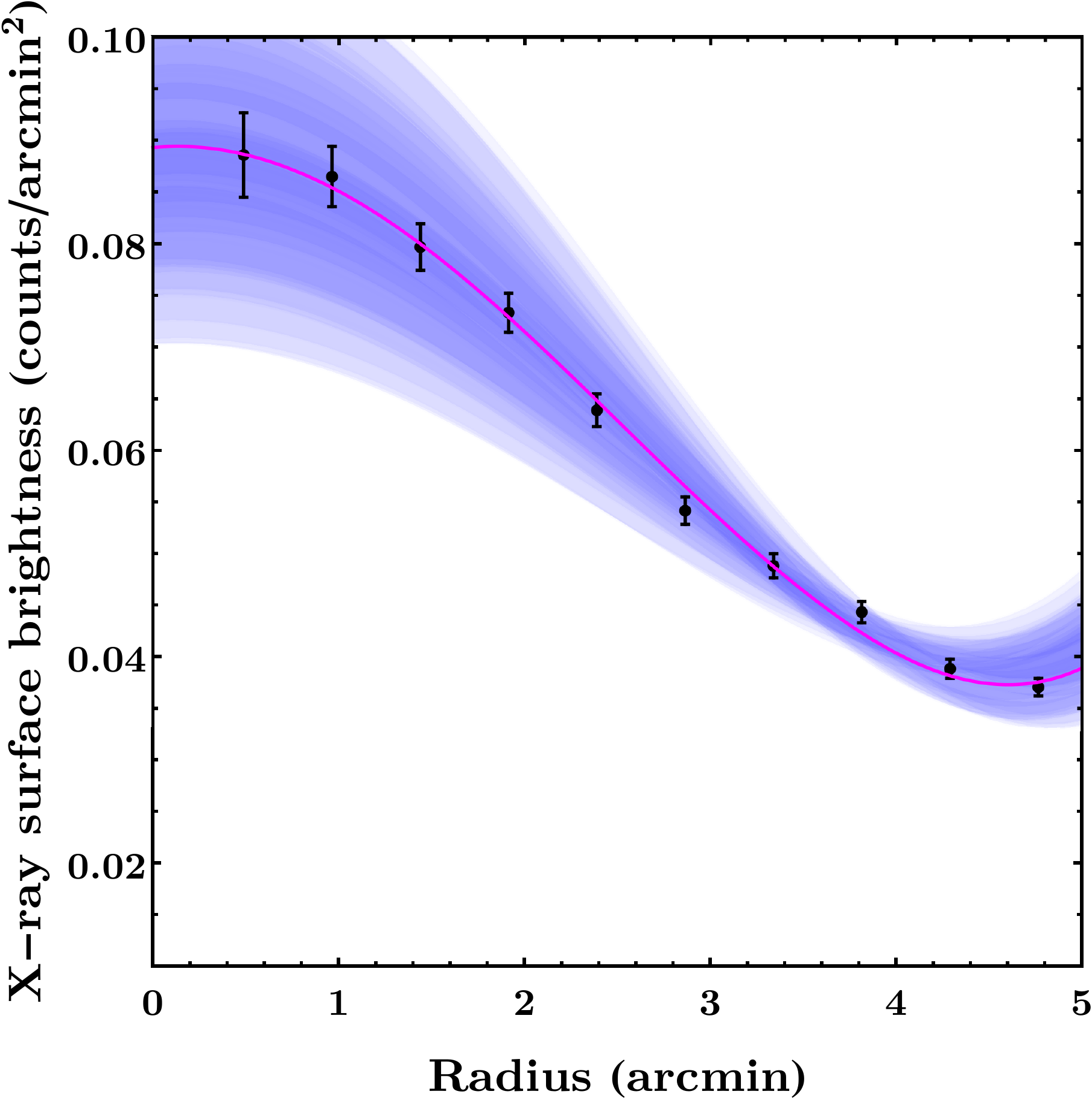}
 		\includegraphics[width=0.48\textwidth]{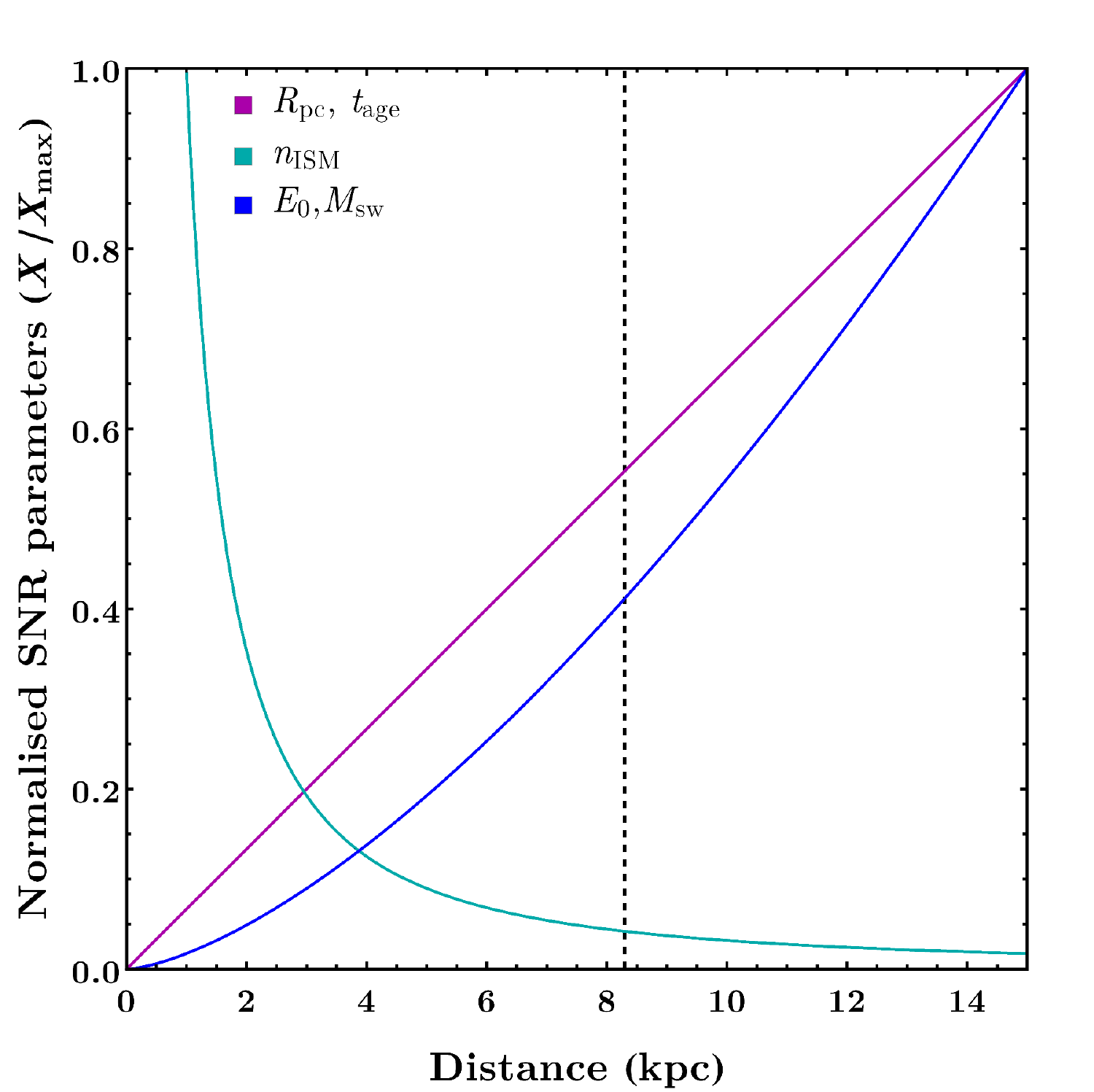}
 		\caption{\textit{Left panel:} Comparision of the measured X-ray surface brightness of G346$-$0.6 with the \citet{1991ApJ...373..543W} similarity solution that best reproduce the observed surface brightness of the remnant.  Here the black data points are from the MOS1 observation of the remnant, the solid magenta line corresponds to the model using the input parameters ($\tau$,  $\mathcal{C}$, $kT_{s}$) = (96, 260, 0.16 keV), while the blue shaded region corresponds to the 68\% uncertainty band for these parameters. \textit{Right panel:} Derived parameters for G346$-$0.2 using the \citet{1991ApJ...373..543W} model. Here we plot the normalised curves for radius ($R_{\rm pc}$), age ($t_{\rm age}$), density of the ISM ($n_{\rm ISM}$), explosion energy ($E_{0}$) and swept up mass ($M_{\rm sw}$), where each curve is divided by $X_{\rm max}$ which is the maximum value of each parameter $X$ derived assuming a distance between $1-15$ kpc. These are derived from Equations \ref{aged}--\ref{vele} and those referenced in the text.  The $X_{\rm max}$ values for $R_{\rm pc}$,  $t_{\rm age}$, $n_{\rm ISM}$, $E_{0}$ and $M_{\rm sw}$ respectively are 22 pc, 7610 kyr, 13 cm$^{-2}$, $1.6\times10^{51}$ erg, and 1284 M$_{\odot}$. The dashed black line corresponds to the values derived assuming a distance of 8.3 kpc which is used throughout of paper.
 			\label{modelling}}
 	\end{center}
 \end{figure*}

In Figure \ref{posteriors} we plot the likelihood functions for $kT_{s}$, $\mathcal{C}$ and $\tau$ that we obtained from our analysis. Here the shaded regions show the 1$\sigma$ confidence intervals for each parameter.  In Figure \ref{modelling} (left) we have plotted the model ($\tau$,  $\mathcal{C}$, $kT_{s}$) = ($94^{+23}_{-36}, 260^{+63}_{-96}, 0.16\pm0.02$ keV) and its uncertainty which best reproduces the surface brightness profile (black data points) and global X-ray temperature of G346.4$-$0.2.

\subsection{The inferred properties of the remnant}\label{properties}

Using these parameters we infer the properties of the remnant using the following equations. The ISM density, $n_{\rm ism}$, is derived using Equation (25) from \citet{1991ApJ...373..543W}, while the age of the remnant is calculated using
\begin{equation}\label{aged}
t_{\rm snr}=\left[\frac{16\pi \mu m_{\rm H} n_{\rm ism}}{25(\gamma+1)\kappa E_{0}}\right]^{1/2}R_{\rm snr}^{5/2} \rm \,\,   yr, 
\end{equation}
which is derived from Equation (8) of \citet{1991ApJ...373..543W}. Here $R_{\rm snr}$ is the radius of the remnant, $E_{0}$ is the 
supernova explosion energy, $\mu=0.604$ is the mean molecular weight, $m_{\rm H}$ is the mass of a hydrogen atom, $\gamma$ is the adiabatic index of the surrounding material, and $\kappa$ is the ratio of thermal and kinetic energy of the remnant as inferred from the similarity solutions of \citet{1991ApJ...373..543W}. $\kappa$ is a function of $\tau$ and $\mathcal{C}$.
 
The explosion energy of the remnant is calculated using 
\begin{equation}\label{enee}
E_{0}=\frac{2(\gamma+1)\pi kT_{s}n_{\rm ism} R_{\rm snr}^{3}}{(\gamma -1)\kappa} \rm \, erg.
\end{equation}
 The shock velocity is calculated using 
\begin{equation}\label{vele}
v_{s}=\frac{2}{5}\left[ \frac{25(\gamma+1)\kappa
	E_{0}}{16\pi\mu m_{\rm H} n_{\rm ism}}\right]^{1/5}t_{\rm snr}^{-3/5} \rm \,  km\, s^{-1}.
\end{equation}
which is derived from Equation (6) of \citep{1991ApJ...373..543W}, while the total mass of X-ray emitting gas is derived from Equation (26) of \citet{1991ApJ...373..543W}. 

In Figure \ref{modelling} right, we plot the  properties of the remnant as a function of distance. Assuming a distance of 8.3 kpc we derive a shock radius of $\sim$12 pc, an explosion energy of $\sim7\times10^{50}$ erg, an age of $\sim$4200 years, a shock velocity of $\sim$1120 km s$^{-1}$, a gas density of the ISM just outside the shock of $\sim$0.52 cm$^{-3}$ and a total swept up mass of $\sim$53 M$_{\odot}$. We find that this model ($\tau$,  $\mathcal{C}$, $kT_{s}$) = (94, 260, 0.16 keV), is able to reproduce the X-ray surface brightness profile (solid magenta line in Figure \ref{modelling} left) and global X-ray temperature of G346.6$-$0.6 quite well.

Similar to previous studies of other MM SNRs using the \citet{1991ApJ...373..543W} model \citep[e.g.,][]{2002ApJ...564..284S, 2004ApJ...616..885C}, the evaporation timescales ($\tau$) of the cold dense cloudlets of material inferred by the best fit model are quite long compared to the age of the SNR ($>$50 $t_{\rm snr}$). The age inferred from our study is lower than that derived by \citet{2011MNRAS.415..301S}. These authors estimated $\sim$11\,kyr for G346.6--0.2 using the ionisation timescale and electron density determined from their modelling. However, we note that they underestimated the X-ray emitting volume of the remnant, which can lead to a larger electron density and thus a larger age.

The derived explosion energy ($\sim7\times10^{50}$ erg) from this model is somewhat low compared to the canonical value for SNRs ($\sim10^{51}$ erg). However, it has been shown that the \citet{1991ApJ...373..543W} model underestimates the explosion energy of SNRs \citep[e.g.,][]{1997ApJ...488..781H}. 

The shock velocity inferred from this analysis is much higher than the typical velocities found in MM SNRs, which of the order of 100\,km s$^{-1}$ \citep[see e.g.,][and reference therewithin]{2015SSRv..188..187S}. In addition, the electron temperature derived from the X-ray analysis ($kT=0.2$ keV) is much lower than one would expect for a shock velocity this fast assuming $kT = (3/16) \mu m v_{s}^2$. One possible explanation for this is that the high velocity and the low temperature of the thermal plasma could result if the recombining plasma originates from a fast shock that has broken through dense material and is now expanding into a low density environment. In Section \ref{rrc} we find that the overionised nature of the plasma most likely arises from adiabatic cooling that is produced in this type of scenario, making it possible that these properties arise from the unique environment of the remnant. 

Another possibility is that the SNR is expanding into a clumpy environment much like that presented by \citet{1991ApJ...373..543W}. As the shock passes through the dense clumpy material and into the lower density interclump medium, this causes rapid cooling producing an overionised plasma, while the high shock velocity could arise from the shock front travelling through the lower density (compared to the dense cloudlets) interclump medium. However, more detailed modelling would be required to shed light on this issue, which is the beyond the scope of this paper and we leave for future work. 

Interestingly though, this model estimates that the amount of swept-up material is  $\sim$53 M$_{\odot}$, which is quite small compared to other MM SNRs \citep[e.g.,][]{2002ApJ...564..284S, 2015ApJ...810...43A} which have swept up $\sim$100$M_{\odot}$. However these remnants are usually much older, allowing them to sweep up significantly more material.

\section{Nature of the thermal X-ray emission}
\subsection{Origin of the sub-solar abundances}
Our spectral analysis indicates sub-solar abundances of Mg, Si, and S in
SNR G346.4$-$0.2, similar to the results derived using \emph{Suzaku}
\citep{2011MNRAS.415..301S,2013PASJ...65....6Y}.
It was suggested that the majority of these elements are not detected because
they have not yet been heated by the supernova reverse shock
\citep{2011MNRAS.415..301S}. However, we argue that this is unlikely, since the sub-solar abundances are found throughout the remnant and the large mass of X-ray emitting material implies that the X-ray emission arises predominantly from shocked ISM, rather than ejecta.

As
\textit{Spitzer} MIPS observations show evidence for a significant amount of
dust associated with the remnant \citep{2006AJ....131.1479R}, one possible
explanation for the subsolar abundances of at least Mg and Si is that these elements have condensed
onto grains, lowering their gas-phase abundance. This is also supported by
\textit{Spitzer} IRS measurements that show strong H$_{2}$ lines from the
interaction of the SNR's shock with dense gas \citep{2009ApJ...694.1266H},
where condensation onto dust grains is likely. However, as S is not a refractory element, this explanation therefore cannot explain its underabundance. As a significant non-thermal X-ray component (whose potential origin in discussed in Section \ref{nth}) has also been detected in our analysis, another possibility is that the synchrotron continuum is dominating the observed X-ray spectrum. \citet{2001ApJ...550..334B} determined that a strong synchrotron continuum can cause X-ray lines to appear to be much weaker than that of a solar abundance plasma. 

\subsection{Origin of the recombining plasma}\label{rrc}
A useful way to characterise the ionisation state of an SNR
plasma is to compare the ionisation temperature $kT_{Z}$, which describes the
extent that ions are stripped of their electrons, with the current
electron temperature $kT_{e}$ of the plasma. Here, $kT_{e}$ is derived from the continuum while $kT_{Z}$ is derived from line ratios \citep{2002ApJ...572..897K, 2005ApJ...631..935K}. When $kT_{Z} < kT_{e}$ or $kT_{Z} > kT_{e}$  the
plasma is in a non-equilibrium ionisation state
\citep{1977PASJ...29..813I}, while $kT_{Z}=kT_{e}$ implies that the plasma
is in CIE. Non-equilibrium ionisation states are most commonly seen in young
SNRs in which shocks produce an ionising plasma that reaches collisional
equilibrium after a $10^{4-5}$ years \citep{2005ApJ...631..935K, 2010ApJ...718..583S}. However, observations of a number of MM SNRs using
\emph{ASCA} \citep{2002ApJ...572..897K,2005ApJ...631..935K}, which was later
confirmed by \textit{Suzaku}, determined
that the thermal plasma of these remnants exhibit evidence of recombination,
where $kT_{Z}>kT_{e}$ \citep[e.g.,][]{2009ApJ...705L...6Y}.
Evidence of rapid electron cooling in the spectra of
SNRs appears in the form of radiative recombination continuum or excess
emission near the K$\alpha$ lines of He-like elements. This rapid cooling can
arise from either rapid cooling of electrons due to the interaction of the
hot ejecta with the cold, dense surrounding environment
\citep{1999ApJ...524..179C}, or through adiabatic expansion that can occur
when the shockfront of an SNR expands through a dense circumstellar material
into a low density environment \citep{1989MNRAS.236..885I}.

To determine the origin of this rapid electron cooling, we can calculate the
timescale of each model. The timescale for thermal conduction is given by \citep{1962pfig.book.....S}:
\begin{equation}
t_{\rm cond} \sim  k_{B} n_{e} l_{T}^{2}/\mathcal{K} ~,
\end{equation}
where $l_T$
is the length of the thermal temperature gradient, $k_B$ is Boltzmann's
constant, $\mathcal{K}$ is the thermal conductivity for a hydrogen plasma, and $n_{e}$ is the electron density\footnote{To derive $n_{e}$ we use $n^2 =4\times 10^{14}  \pi K d^{2} f^{-1}
V^{-1}$, where $K$ is the normalisation of our global spectrum, $f$ is the filling factor, $n_e = 1.2n_{\rm H}$, $n \approx
1.1n_{\rm H}$ assuming \citet{2000ApJ...542..914W} abundances and $V$ is the volume of our extraction region. To derive the volume, we assume a filled ellipsoid with semi-major radius of 3.4\arcmin
(or $8.3\,d_{8.3}$\,pc) and a semi-minor radius of 3.0\arcmin (or
$7.3\,d_{8.3}$\,pc), which corresponds to a volume of
$V=5.4\times10^{58}\,f\,d_{8.3}^{3}$ cm$^{3}$.}. We take the length of the semi-major axis of the region we used to extract the global X-ray spectrum as the length scale, i.e., $2.5\times10^{19}d_{8.3}$\,cm. Using the temperature and $n_e$
associated with the global X-ray spectrum, we derive the thermal conduction
timescale of $\sim$500\,kyr. Much like other MM SNRs such as IC443 \citep{2009ApJ...705L...6Y}, and MSH 11$-$61A \citep{2015ApJ...810...43A}, the timescale derived using thermal conduction is significantly larger than the age of the
remnant we derived in Section \ref{ages}, indicating that efficient thermal conduction
is most likely not responsible for the rapid electron cooling.

For adiabatic cooling, the timescale can be estimated using the ionisation timescale of the plasma $n_e = \tau_{i} t$, 
where $\tau_{i}$ is the ionisation timescale of the plasma as derived from
modelling the global X-ray spectrum. This gives us an adiabatic cooling timescale of
$\sim12$\,kyr, which is comparable to the age of the remnant as derived in
Section~\ref{ages}. This implies that the origin of the recombining plasma is most likely adiabatic cooling \citep[e.g.,][]{2013ApJ...777..145L}.

 \subsection{Supernova type}\label{progenitor}
 The supernova type of G346.6--0.2 is currently not well
 identified. On one hand, it was suggested that G346.6--0.2 could arise from a Type Ia SN explosion based on the relative
 abundance of Fe compared to Si \citep{2011MNRAS.415..301S}. On the other hand, \emph{Spitzer} detection of spectral lines associated with shocked H$_2$ indicate that it could be a core collapse (CC) event \citep{2009ApJ...694.1266H, 2014AJ....147...55P}.
 There are a number of way to shed light on the progenitor of an SNR. This
 includes comparing chemical abundances such as the O/Fe ratio to the values
 predicted by Type Ia and CC SN models \citep[e.g.,][]{1999ApJS..125..439I}
 since Type Ia SNe produce significantly more Fe than CC SNe, while CC SNe
 produce a large amount of O compared to Type Ia SNe. One can also determine
 associations with nearby molecular clouds or by analysing the asymmetry of the
 remnants morphology, with CC SNe being more asymmetric that Type Ia SNRs
 \citep{2011ApJ...732..114L, 2009ApJ...706L.106L}. 
 
 Unfortunately, due to the relatively large absorption in the
 direction of G346.6--0.2, the poor statistics above 5 keV and the possibility that the X-ray emission arises from shocked ISM, we are unable to
 use chemical abundances derived from our X-ray spectra to classify whether this remnant is a Type Ia or CC.  However, we argue that G346.6--0.2 arises
 from a massive progenitor that underwent CC SN, based on its association with a dense molecular cloud, as well as the detection of infrared emission
 from shocked molecular gas \citep{2006AJ....131.1479R, 2009ApJ...694.1266H,
 	2011ApJ...742....7A}, the highly asymmetric nature of its X-ray morphology,
 and the fact that this remnant is located in the Galactic plane.

\section{Nature of the hard X-ray tail component}\label{nth} 
The detection of non-thermal X-rays could originate from a number
of possibilities, including an undetected pulsar or pulsar wind nebula (PWN), contamination from a
nearby source or a population of sources, Galactic Ridge X-ray emission, or a population of relativistic
particles accelerated by the SNR shock-front.

\subsection{Galactic Ridge X-ray emission?}

G346.6--0.2 is located in the inner part of the Galactic disk which is
dominated by strong X-ray emission from the Galactic Ridge X-ray Emission (GRXE).  It has been shown in both deep \emph{Chandra} \citep[e.g.,][]{2005ApJ...635..214E, 2009Natur.458.1142R} and \emph{Suzaku} \citep[e.g.,][]{2012ApJ...753..129Y,2013PASJ...65...19U} studies that the majority of this emission can be resolved into faint X-ray emitting stellar coronae and accreting white dwarf binaries. \citet{1997ApJ...491..638K} showed that in the 0.5--10.0 keV energy range, the GRXE is best described by an optically-thin thermal plasma model with a low and high temperature component of $\sim$1.0 keV and $\sim$6 keV respectively. This was later confirmed using \emph{Chandra} \citep[e.g.,][]{2005ApJ...635..214E} and \emph{Suzaku} \citep[e.g.,][]{2009PASJ...61..751R, 2012ApJ...753..129Y, 2013PASJ...65...19U}. \citet{2012ApJ...753..129Y} attempted to decompose the high energy ($\sim$6 keV) component of the GRXE into its various discrete source contributions. These authors found that this high energy component is best described by a thermal plasma in CIE with a temperature of $\sim1.2-1.7$ keV arising from coronal X-ray sources and a spectral component that arises from accreting white dwarfs with a mass $\sim0.7M_{\odot}$.

The GRXE can be a major background contribution to the X-ray spectrum of
 diffuse sources, particularly for energies greater than 5 keV \citep[e.g.,][]{2013PASJ...65....6Y}, or if the
 statistics of a source's X-ray spectrum are poor. As such to test whether the hard X-ray tail seen in our fits arises from the high temperature component of the GRXE we fit the global X-ray spectrum in two ways.

First we fit the global X-ray spectrum with an absorbed VRNEI plus a APEC model in which the temperature is set free. As discussed in more detail in Section \ref{spec}, we find that we can produce quite a good fit (reduced $\chi^2=0.96$) to our X-ray spectrum with the temperature of the APEC model $\sim1.9$ keV. While this value is much lower than 6 keV derived in previous studies \citep[][]{1997ApJ...491..638K, 2005ApJ...635..214E, 2009PASJ...61..751R}, it is comparable to the softer, coronal X-ray source contribution of the high temperature component of the GRXE derived by \citet{2012ApJ...753..129Y}, or the low temperature component of the GRXE derived by \citet{2013PASJ...65...19U}.

Second, we fit the global X-ray spectrum using an absorbed VRNEI plus a APEC model in which we fix the temperature, and abundance parameters derived by \citet{1997ApJ...491..638K} or \citet{2013PASJ...65...19U} for the the GRXE.  Again, we find that this produces a similar result ($\chi^{2}_{r}  = 0.99$) compared to our model listed in Table \ref{fits} and \ref{fitsapec}. As a consequence, we cannot rule out that the additional hard component required by our fits arises from the GRXE.

Due to the relatively small size of the G346.6--0.2 (diameter $\sim0.1^{\circ}$ in diameter), we do not expect the GRXE to vary significantly across the remnant considering the GRXE varies on scale heights of 0.5$^{\circ}$ latitudinally and 3$^{\circ}$ longitudinally \citep{1997ApJ...491..638K}. As a consequence we would expect that if the GRXE is responsible for the observed hard X-ray tail seen in Figure \ref{xrayspectrum} that our best fit absorbed VRNEI+POWERLAW (or VRNEI+APEC) models would produce similar $\Gamma$ (or temperatures) values. However, from Table \ref{fits} and \ref{fitsapec} one can see that we do see quite a large variation in best fit photon index (temperature) we obtain, while region 2 does not require an additional powerlaw or APEC component.

In addition, \citet{2013PASJ...65....6Y} carefully took into account the variation in the GRXE with scale height when modelling the X-ray spectrum from G346.6--0.2 using \textit{Suzaku} by using different background regions located at different positions along their field of view for their fit. Nonetheless, they were unable to obtain a good fit with a $\chi^{2}_{r} < 1.2-1.4$ implying that the spectrum might require an additional component. However, deeper observation using \textit{XMM} and/or \textit{NuSTAR} would be required to constrain the higher energy component of this remnant and to confirm or rule out the possibility of the GRXE contributing to the observed X-ray spectrum.

 \begin{figure*}[t!]
	\begin{center}
		\includegraphics[width=0.50\textwidth]{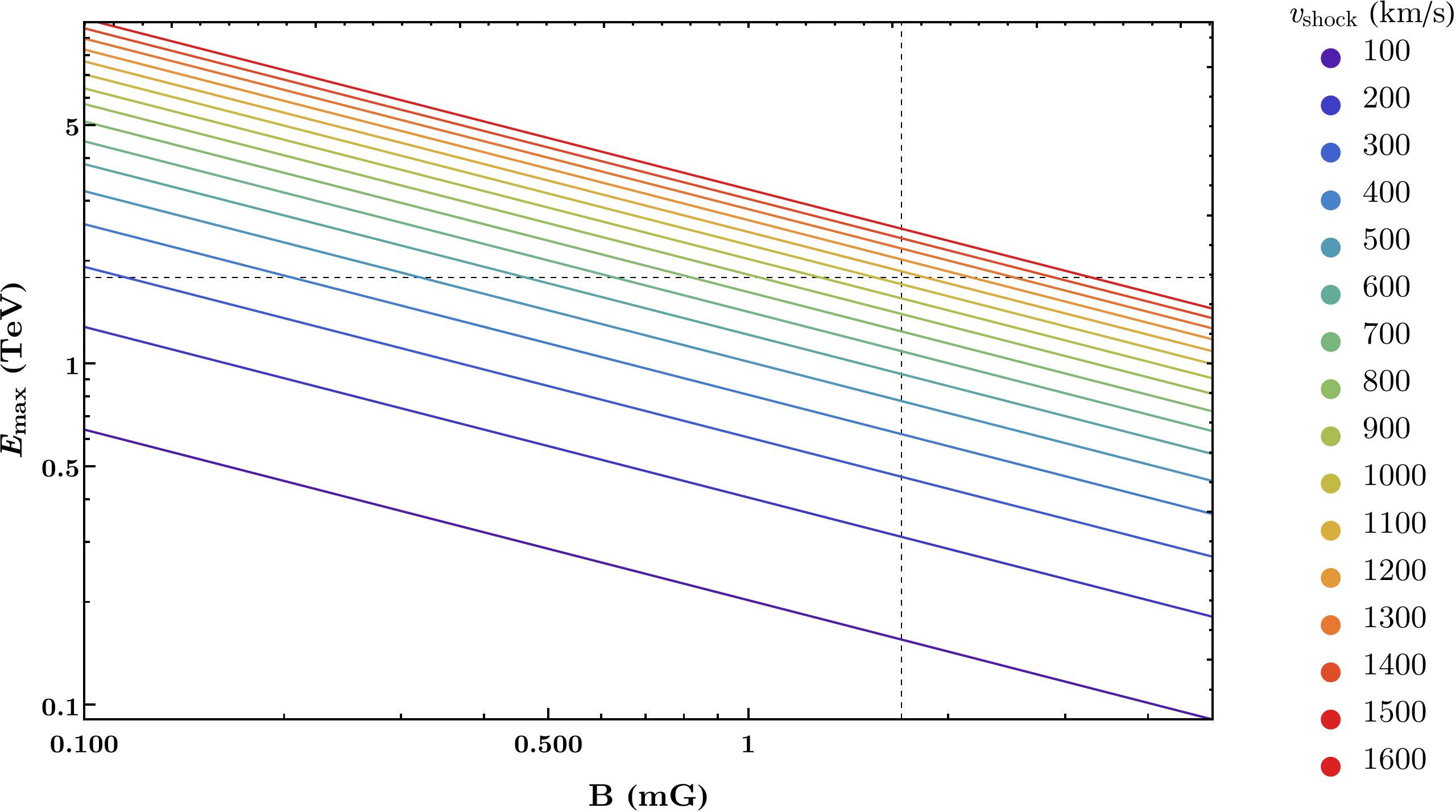}
		\includegraphics[width=0.48\textwidth]{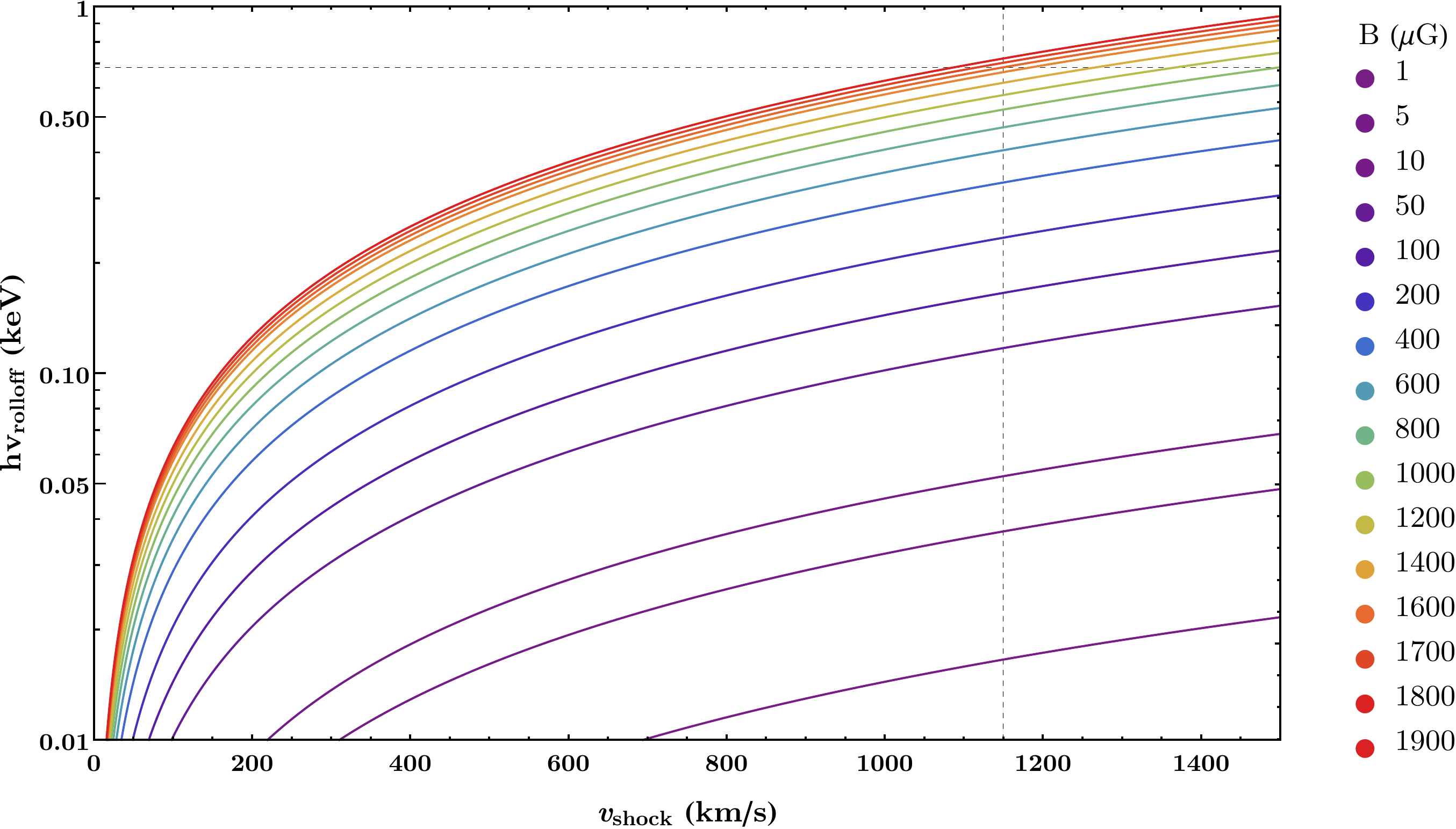}
		\caption{\textit{Left panel:} The maximum energy (Eq.~\ref{maxen}) of accelerated electrons ($E_{max}$) as a function of magnetic field ($B$) for various shock velocity ($v_{shock}$) plotted on a log-log plot. Here we have assumed  $\eta=1$ and  $(\chi-1/4)/\chi^2=1$.  The dashed lines correspond to the $E_{max}=1.8$ TeV assuming the $v_{shock}$ value inferred in Section \ref{properties} and $B=1.7$\,mG as derived from the VLA polarimetric observations of the OH masers toward G346.6--0.2 \citep{1998AJ....116.1323K} . \textit{Right panel:} Rolloff energy ($h\nu_{\rm rolloff}$) as a function of  $v_{shock}$ for different $B$ (Eq.~\ref{roll}). The dashed lines correspond to $h\nu_{\rm rolloff} = 0.7$ keV assuming the $v_{shock}$ value inferred in Section \ref{properties}, $B=1.7$\,mG, $\eta=1$ and $(\chi-1/4)/\chi^2=1$.
			\label{emaxnu}}
	\end{center}
\end{figure*}

\subsection{Unidentified pulsar/pulsar wind nebula?}
The core collapse of a massive star can lead to the formation of a neutron star. Neutron stars are rapidly rotating and have strong magnetic fields, forming a highly relativistic wind of particles. The particles in this pulsar wind interact with surrounding photon and magnetic fields, emitting both synchrotron and inverse Compton radiation which are observed as a PWN. The emission from a PWN can be well described by a power-law spectrum with $\Gamma\sim$0.5--2.0 \citep[see e.g.,][]{2006ARA&A..44...17G}, which is consistent with the photon index derived using our power-law model. Additionally, based on the global X-ray spectrum, this component has an average unabsorbed flux of $8.5\times10^{-13}$\,erg\,cm$^{-2}$\,s$^{-1}$ over the 0.5--7\,keV energy band. This corresponds to a luminosity of $7.3\times10^{33}d_{8.3}$\,erg\, s$^{-1}$, which is comparable to the observed X-ray luminosity of a large number of PWN detected in the X-ray energy band \citep{2008AIPC..983..171K}. 

As G346.6--0.2 most likely formed from a CC SN (see Section~\ref{progenitor}), it is possible that the observed non-thermal emission arises from an unidentified pulsar and its nebula. As discussed in more detail in Section~\ref{neutronstarsearch}, we search for a neutron star candidate within a circular radius defined if one assumes that the supernova explosion which created G346.6--0.2 produced a neutron star with a kick velocity of $\sim100-500$ km s$^{-1}$. We find one potential candidate within the assumed distance from the remnant centre (source 8 in Figure \ref{pointsrc}).

As its emission is relatively soft in nature (HR$=-0.2$), this source could potentially be a young thermally emitting neutron star, similar to Cas A \citep[e.g.,][]{2009ApJ...703..910P} or G350.1--0.3 \citep[][]{2011ApJ...731...70L}. However, unlike the two cases above, our neutron star candidate exhibits some hard ($>2$ keV) X-ray emission. Therefore, it is possible that this source harbours an extended PWN which is responsible for the hard X-ray component detected in our analysis. By studying the exposure-corrected hard ($>3$ keV) X-ray image of the remnant we find that this emission is concentrated around the position of this neutron star candidate, and shows faint extended emission that extends from the position of the point source.

\subsection{Particles accelerated by the shock-front?}

\subsubsection{IC Scattering or Non-thermal Bremsstrahlung?}
Non-thermal X-ray emission in SNRs can arise from three main sources of emisson: inverse Compton (IC) scattering, non-thermal bremsstrahlung, or synchrotron emission from shock accelerated particles. For energies less than 10\,keV and magnetic field between 5-500$\mu$G which are typically found in SNRs, IC scattering is not thought to be the dominant mechanism producing non-thermal X-ray emission in SNRs \citep{2012A&ARv..20...49V}. Thus we do not consider the case that IC scattering is responsible for the observed power-law component.

Non-thermal bremsstrahlung arises from non-relativistic electrons which lose their energy via Coulomb interactions. This causes the thermalisation of the low energy tail of electrons, producing a relatively steep spectral index ($\Gamma \sim s-1<$1.5, where $s$ is the index of the electron population and $\Gamma$ is the photon index) for any non-thermal emission detected. However, \citet{2008A&A...486..837V} determined that non-thermal bremsstrahlung dominates for short ionisation timescales ($n_{e}t \sim 10^{10}$ cm$^{-3}$ s) and thus is only expected to be found in a narrow region close to the shock front. For remnants like G346.6--0.2, which have long ionisation timescales $>$10$^{11}$ cm$^{-3}$ s, the non-thermal bremsstrahlung component is dominated by X-ray continuum, and/or synchrotron X-ray emission  \citep{2008A&A...486..837V, 2012A&ARv..20...49V}. As a consequence it is expected that the non-thermal emission detected in the soft X-ray band most likely arises from another mechanism. 

\subsubsection{Synchrotron emission?}

Another possible scenario is that the hard X-ray tail arises from synchrotron radiation produced by a population of electrons accelerated by the SNR shock front. Non-thermal emission is usually detected in young ($< 1$ kyr)
SNRs which have fast moving shocks with velocities $> 2000$\,km\,s\,$^{-1}$
and magnetic fields between 50--250\,$\mu$G \citep{1999A&A...351..330A, 2006AdSpR..37.1902B}. Assuming that the particle acceleration is limited by synchrotron loss, the maximum energy of the underlying electron population can be estimated using \citep{1999A&A...351..330A, 2012A&ARv..20...49V}:
 	\begin{eqnarray}\label{maxen}
 	E_{\rm max}  \sim 32 \eta^{-\frac{1}{2} }\left( \frac{B}{100 \rm \mu G}\right)^{-\frac{1}{2}} \left( \frac{v_{s}}{5000 \rm \, km \, \rm s^{-1}}\right) \left(\frac{\chi-\frac{1}{4}}{\chi{2}}\right)^{\frac{1}{2}}\,\mbox{TeV}.
 	\end{eqnarray}
Here $B$ is the magnetic field, $v_{s}$ is the shock velocity, $\eta$ is the particle acceleration efficiency and $\chi$ is the compression ratio. Under these conditions, young SNR can accelerate electrons up to energies of
$10-100$ TeV \citep{2012A&ARv..20...49V}. 

Older SNRs like MM SNRs usually have shock velocities that are effectively too slow to
accelerate the electrons to the energies needed to produced X-ray synchrotron
emission, assuming a magnetic field of 10--100s of $\mu$G. However, using VLA polarimetric observations of the OH masers
toward G346.6--0.2, \citet{1998AJ....116.1323K} found a
magnetic field of $B=1.7$\,mG for G346.6--0.2 using Zeeman splitting. 
The lines inferred from molecular line measurements are produced from shocks being driven into cold, dense material. These conditions do not necessarily represent the properties of the X-ray emitting region of the remnant, and thus the magnetic fields derived from these line measurements could be much higher than what is found in the bulk of the cloud material. As such, we take $B=1.7$\,mG as an \textit{upper limit} to the magnetic field across the remnant. This value implies that the shock
velocity derived in Section \ref{properties} for G346$-$0.2 does not need to be as high as those seen in other X-ray synchrotron SNRs (see Eq.~\ref{maxen}). 

From Equation \ref{maxen} we can derive an \textit{upperlimit} to the maximum energy of the accelerated electrons. Assuming the shock velocity derived in Section \ref{properties} and the magnetic field of G346.6$-$0.2 derived using Zeeman splitting, the maximum energy of the underlying electron population for G346.6$-$0.2 is $E_{\rm max}\sim \eta^{-1/2}((\chi-\frac{1}{4})/\chi^{2})^{-1/2}$ TeV. This value is slightly lower than that of other X-ray synchrotron emitting SNRs such as 5--12 TeV for Tycho \citep{2015ApJ...814..132L} and 5 TeV \citep{2004ApJ...602..271L} for RX J1713.  
 
As the non-thermal X-ray and radio emitting regions of G346.6--0.2 are not correlated (see Figure \ref{xrayimage}), it is likely that there are different electron populations producing the radio and X-ray synchrotron emission. Thus, unlike other studies such as \citet{1999ApJ...525..368R} which assume that the radio and X-ray emission arises from the same particle population, we are unable to fit the global X-ray spectrum using the XSPEC model \textit{srcut} to characterise the underlying particle energy distribution. However, even though we are unable to do this directly, assuming that the particle acceleration is limited by synchrotron loss, we can estimate an \textit{upperlimit} to the rolloff frequency $h\nu_{\rm rolloff}$ of the underlying particle population using \citep{1999ApJ...525..368R}:
 \begin{eqnarray}\label{roll}
h\nu_{\rm rolloff} \sim \rm 5\times10^{15}\left(\frac{\textit{B}}{10 \, \mu G}\right) \left(\frac{\textit{E}_{\rm max}}{10 \, TeV} \right)\,keV
\end{eqnarray}
Using the shock velocity inferred in Section \ref{properties} and $E_{\rm max}$ derived in Equation \ref{maxen}, we estimate $h\nu_{\rm rolloff} \sim 0.7 \eta^{-1/2} (\chi-\frac{1}{4})/\chi^{2} $ keV. This is comparable to that derived by \citet{1999ApJ...525..368R} for G346.6--0.6 using \emph{ASCA}, as well as other SNRs with non-thermal X-ray emission such as Tycho (e.g., \citealt{2002ApJ...581.1101H}). 

In addition to the simple exercise above, we also investigated how $h\nu_{\rm cutoff}$ and $E_{max}$ of the electron population changes for different magnetic field values and shock velocities. From Figure \ref{emaxnu} one can see that as $v_{shock}$ decreases for fixed $B$-field, both $h\nu_{\rm cutoff}$ and $E_{max}$ decrease. Similarly, if we fix $v_{shock}$, and decrease the $B$-field, $h\nu_{\rm cutoff}$ and $E_{max}$ decrease.  Currently, known X-ray synchrotron emitting SNRs such as Tycho and SN1006 have $E_{\rm max}\geq 1$ TeV.   For $v_{shock}$ inferred in Section \ref{properties}, as $B$ decreases  $E_{\rm max}$ will still fall into the range seen in other SNRs that show X-ray synchrotron emission. For a fixed $B$, $v_{shock}$ would have to fall below $\sim 400$ km s$^{-1}$ before $E_{\rm max}\leq 1$ TeV. As we are unable to directly determine the cut-off energy from modeling the X-ray spectrum, and the VLA measured $B=1.7$\,mG for G346.6--0.2 represents an \textit{upperlimit} only, we suggest that $h\nu_{\rm rolloff}$ and  $E_{\rm max}$ inferred from this study represents an upperlimit only.  Further observations of this remnant to search for X-ray synchrotron filaments (e.g., \citealt{2003ApJ...584..758V, 2006A&A...453..387P}) using \emph{Chandra} would allow us to constrain the magnetic field, while \emph{NuSTAR} observations will allow us to confirm and characterise the non-thermal emission of the remnant and thus better constrain the properties of the underlying particle population.

\begin{figure}[t!]
\begin{center}
 		\includegraphics[width=0.48\textwidth]{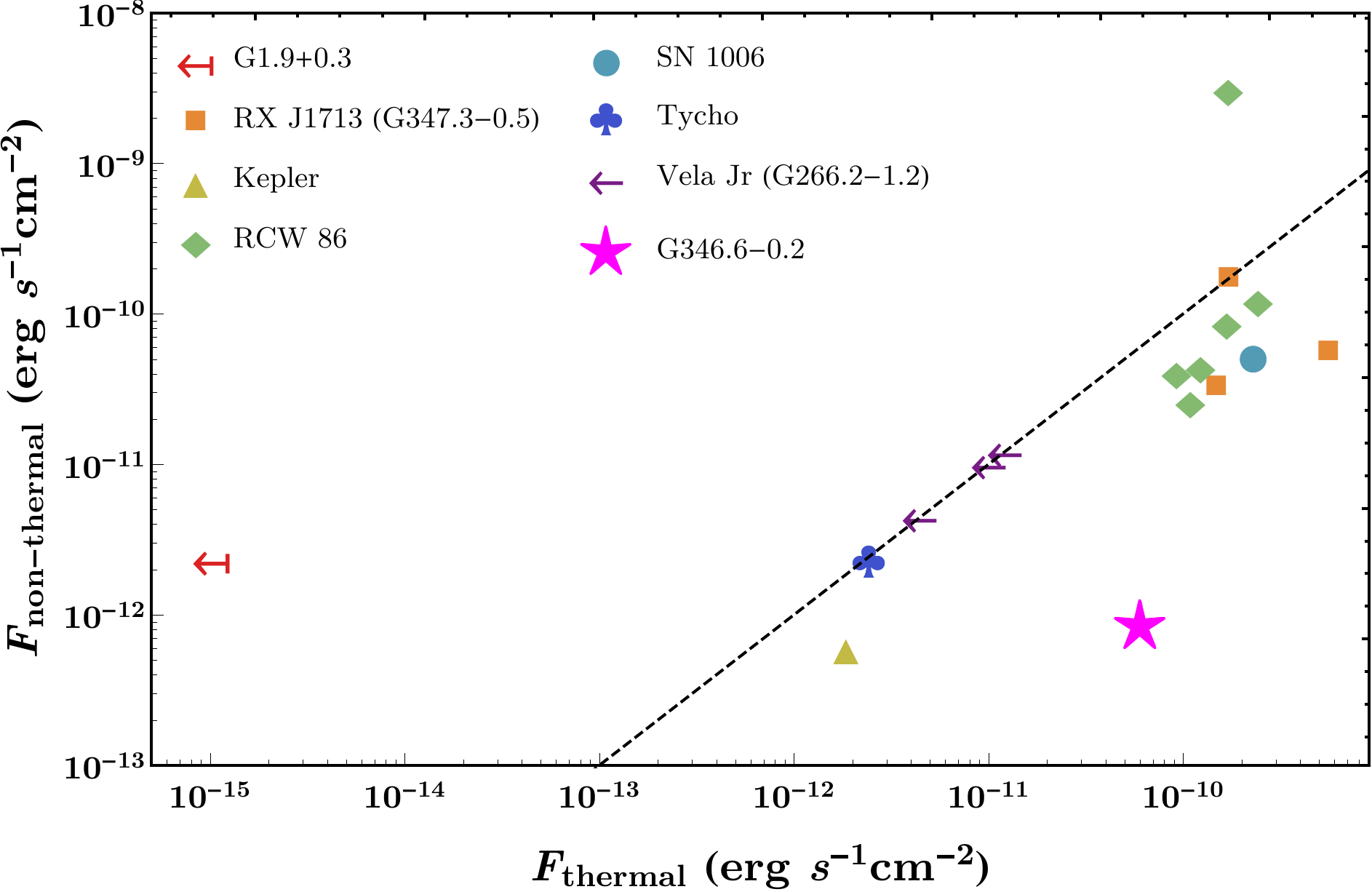}
\caption{The unabsorbed non-thermal flux plotted against the unabsorbed thermal flux of eight well known X-ray synchrotron emitting SNRs as derived using the best fit models of their X-ray emission as presented in the literature. Here the fluxes are derived in the 0.5--7.0 keV energy band. For RX J1713, RCW 86 and Vela Jr, we include the flux measurements derived from multiple regions across the remnant. For G1.9+0.3, and Vela Jr whose emission is significantly dominated by its non-thermal component, we set upper limits to their thermal X-ray emission by either following what is suggested in the literature (for G1.9+0.3), or setting the flux of their thermal emission equal to the flux of their corresponding non-thermal component (for Vela Jr).  Plotted as the magenta star ($\star$) is the unabsorbed thermal and non-thermal flux derived for G346.6--0.2 in this study. Plotted as the black dashed line is when the unabsorbed thermal and non-thermal fluxes are equal.
	\label{comparison}}
\end{center}
\end{figure}

Compared to other X-ray synchrotron emitting SNRs such as Cas A \citep{2001ApJ...552L..39G, 2009PASJ...61.1217M}, G1.9+0.3 \citep{2010ApJ...724L.161B}, Kepler \citep{2004A&A...414..545C}, RCW 86 \citep{2014MNRAS.441.3040B, 2016arXiv161201221T}, RX J1713 \citep{2004ApJ...602..271L, 2015ApJ...814...29K}, SN 1006 \citep{2013ApJ...771...56U}, Tycho \citep{2002ApJ...581.1101H, 2011ApJ...728L..28E}, and Vela Jr \citep{2013A&A...551A...7A}, we find that G346.6--0.2 exhibits a thermal to non-thermal flux ratio that is nearly an order of magnitude larger that what is seen for the other X-ray synchrotron emitting SNRs (see Figure~\ref{comparison}). This is in contrasts to that X-ray synchrotron emitting SNRs listed above which have a thermal to non-thermal flux ratio of $\sim$ 1. If arising from particle's being accelerated by the shock front of the remnant, one possible explanation for the faintness of its non-thermal component compared to its thermal component is the fact G346.6--0.2 is found in a significantly denser environment compared to other X-ray synchrotron SNRs. This could lead to significant cooling that lowers the maximum energy of the underlying electron population.  As a result, fewer electrons can produce the non-thermal emission, reducing the emissivity.

Compared with other MM SNRs, only W49B is known to have shock velocities of the same order as G346.6--0.2 \citep{2007ApJ...654..938K}, but as of writing no synchrotron emission has been detected from this remnant. All other MM SNRs have velocities between 50--200 km s$^{-1}$ which are too slow to produce X-ray synchrotron emission, consequently their X-ray emission is primarily thermal in nature\footnote{We note that using \emph{Suzaku}, \citet{2009PASJ...61S.155K} were able to fit the northwestern X-ray emission of MM SNR G156.2+5.7 using an absorbed VNEI+VNEI+POWERLAW model. They concluded that this power-law component arose from X-ray synchrotron emission from inefficient particle acceleration of a population of electrons by a shock with a velocity of $\sim500$\,km\,s$^{-1}$. However, their fits also imply that an VNEI+VNEI+NEI model is able to reproduce the observed X-ray emission in this region equally well. \citet{2012PASJ...64...61U} reanalysed the \emph{Suzaku} data and found that the power-law component is more likely associated with the cosmic X-ray background. Thus a more detailed study of this object would be required to confirm the presence (or lack) of X-ray synchrotron emission from this remnant.}. Of the MM SNRs in which non-thermal X-ray emission has been detected, such as W28 \citep{2014ApJ...791...87Z}, IC443 \citep{2001A&A...376..248B} and W44 \citep{1996ApJ...464L.165F}, this emission arises from non-thermal bremsstrahlung or a PWN. 

For SNRs found in dense environments, their shocks are usually radiative \citep[see e.g.,][]{2002cosp...34E.970B, 2015SSRv..188..187S}, which can produce a large compression ratio leading to a lower $h\nu_{\rm rolloff}$ and $E_{max}$. However, even if the compression ratio $\chi$ is three to four times higher than that, $h\nu_{\rm rolloff}$ can still easily fall within the X-ray emitting band such that G346.6--0.2 could still possibly accelerate particles to synchrotron emitting energies. In addition, radiative shocks can also produce highly compressed magnetic fields. This can lead to significant magnetic field amplification of ISM magnetic fields, as well as enhanced cosmic-ray electron densities, which can result in strongly enhanced radio emission and to a lesser extent enhanced X-ray synchrotron radiation (see e.g., \citealt{2012A&ARv..20...49V} and references therein). 

For the SNRs which have OH masers, a large fraction of these are MM SNRs. OH masers are preferentially located in regions of dense molecular material that have recently been shocked, and the magnetic fields of the MM SNRs with OH masers as derived from Zeeman splitting range between $0.2-2.2$ mG \citep[e.g.,][]{1997ApJ...489..143C, 2000ApJ...537..875B, 2013ApJ...771...91B}. On average, this is much larger than the magnetic fields found in shell type X-ray synchrotron emitting SNRs which range from 10--100s $\mu$G \citep{2006AdSpR..37.1902B}. The unique properties of G346.6--0.2 as well as the possible combination of radiative shocks and the presence of a dense environment might lead to the production of X-ray synchrotron emission in this remnant. If this power-law component is confirmed to arise from particles being accelerated by the shock-front, this would make G346.6--0.2 an important new object in the class of synchrotron emitting SNRs. However, deep observations of this source are required to confirm the origin of this component.

\section{Searching for a neutron star candidate} \label{neutronstarsearch}

As we discussed in Section~\ref{progenitor}, G346.6--0.2 is likely a remnant
of a CC SN. Detecting an associated pulsar or PWN can directly confirm this scenario. The emission from a neutron star with a PWN will be non-thermal in nature and be best described by a power-law spectrum with $\Gamma\sim$0.5--2.0 \citep[see e.g.,][]{2006ARA&A..44...17G}, while for a young neutron star without a bright PWN, its emission can be described by a soft thermal component \citep[see e.g.,][]{2002nsps.conf..273P}. 

\begin{figure}[t!]
	 		\includegraphics[width=1.1\columnwidth]{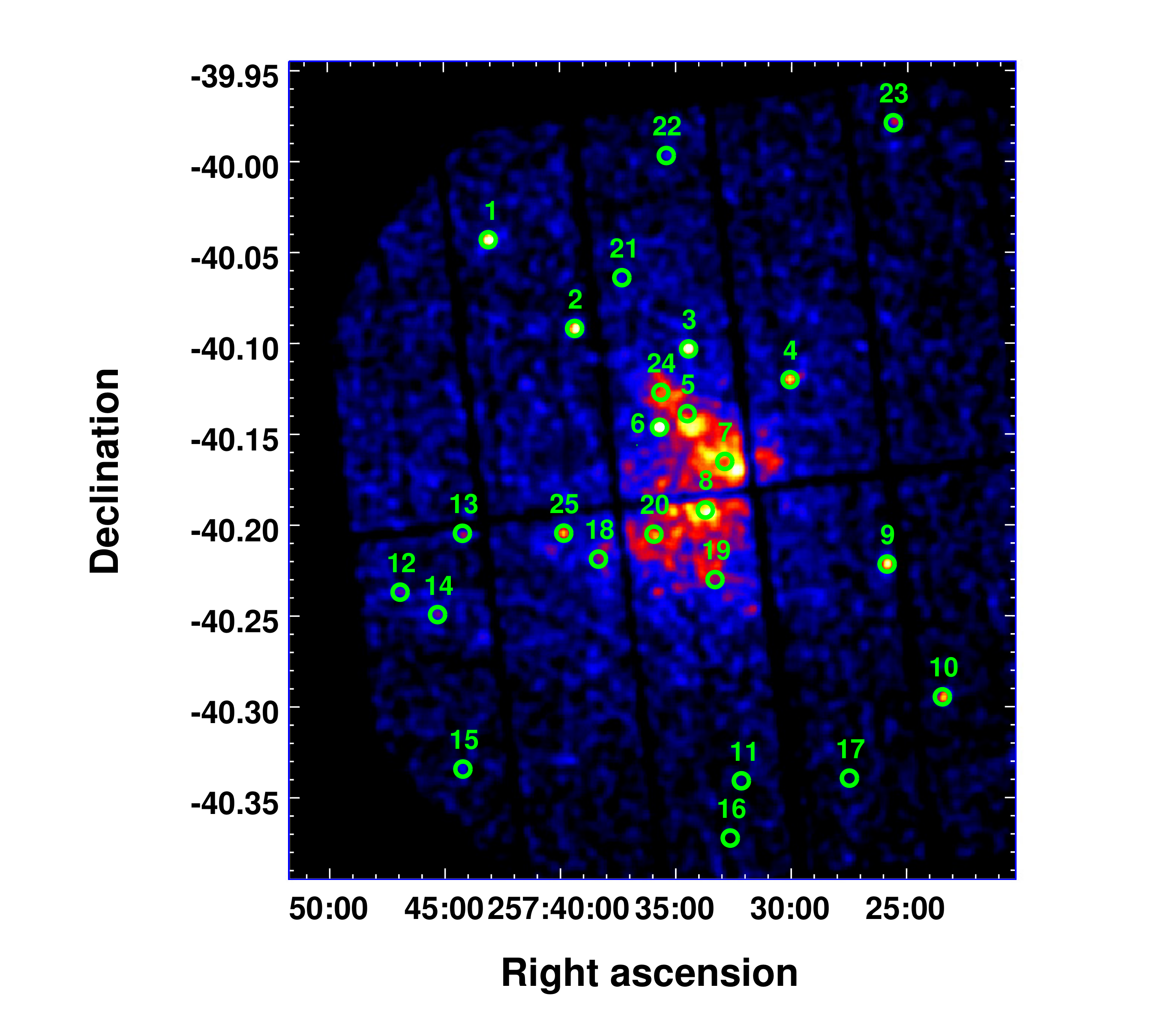}
	\caption{Point sources detected with a likelihood threshold of 30$\sigma$ or
		more using \texttt{edetect\_chain}, overlaid on a 0.5--7\,keV exposure
		corrected PN image of G346.6--0.2 that has been smoothed with a
Gaussian of width 20\arcsec. The sources are labeled 1--24 and their
		properties are listed in Tables~\ref{counterparts} and \ref{ptsrctable}.
 \label{pointsrc}}
\end{figure}

The current telescope
configuration of this observation, is not useful for a timing analysis to
search for pulsations from a possible pulsar candidate.  As a consequence, we attempt to determine the nature of the point sources in the field of view (see Figure~\ref{pointsrc}) by searching for optical/IR counterparts in B2 and R2 USNO-B1 catalogues, while extracting and modelling the X-ray emission from each source using an extraction region with a radius of 15\arcsec (see Table~\ref{counterparts} and \ref{ptsrctable}) centred on the position of the source. For all sources, particularly those that do not have an optical counterpart, we calculate the
HR\footnote{Due to the relatively shallow XMM observation presented in this paper, a more detailed analysis of the point sources similar to \citet{2014ApJS..212...13A} is beyond the scope of this paper.} using ($R_{2-10}-R_{0.2-2}$)/($R_{2-10}+R_{0.2-2}$), where
$R$ is the count rate across either the $0.2-2.0$ keV or $2.0-10.0$ keV energy
band \citep{2010ApJ...725..931M}. Sources with a HR$>$0 are hard X-ray sources, which include objects such as neutron star with a PWN, while those with HR$<$0 are soft X-ray sources such as stars, thermal emission from a SNR (ejecta clumps) or possibly a young neutron star without an X-ray bright PWN. 

Within the field of view, 11 out of the 25 sources detected have no optical or IR counterparts. These are good candidates for a pulsar or a PWN. Here we define a circular radius centred at  $\rm (\alpha,\delta)=(17^{h}10^{m}17^{s}, -40^{\circ}10'59'')$ within which we would expect to find an associated neutron star given a reasonable kick velocity of  $\sim100-500$ km s$^{-1}$ \citep[see e.g.,][]{2002ApJ...568..289A,2005MNRAS.360..974H}. This corresponds to a radius of $\sim$ 0.2\arcmin \, and 0.9\arcmin \, assuming a neutron star velocity of 100 km s$^{-1}$ and 500 km s$^{-1}$ respectively. Assuming a neutron star velocity of 100 km s$^{-1}$, none of the sources we detect fall within this circular radius, while only source 8 falls within the radius expected for a neutron star travelling with a kick velocity of 500 km s$^{-1}$.
	
Source 8 has no optical or IR counterparts and has an HR of $-0.2$ indicating that the emission from this source is relatively soft and possibly thermal in nature. We were able to extract a spectrum from this source, however due to the short exposure time of our observation the uncertainties in our fit parameters are quite large. Regardless of this, we find that the emission from Source 8 can be fit using an absorbed power-law with a column density of $N_{\rm H}=(3.4^{+2}_{-1})\times10^{22}$\,cm$^{-2}$ and a power-law index of $\Gamma=4.4^{+1}_{-2}$. The former is consistent with that of G346.6--0.2 (see Table \ref{fits}), while within uncertainties either a thermal and non-thermal model\footnote{Due to limited statistics, a power-law model with a steep index typically mimics emission arising from a thermal component.} can easily fit the observed X-ray emission. Based on its HR, it is possible that the emission from this source is more thermal in nature and could potentially arise from a young neutron star without a bright PWN. Assuming that this point source is consistent with a neutron star, we find that the 0.5--10\,keV
unabsorbed flux of this source is $\sim9\times10^{-14}$\,erg\,cm$^{-2}$\,s$^{-1}$, which
corresponds to an X-ray luminosity of $\sim8\times10^{32}$\,erg\,s$^{-1}$ at a
distance of 8.3\,kpc. This is similar to the luminosities seen for other X-ray pulsars and their nebulae \citep[see Tables
2--3 in][]{2008AIPC..983..171K}, making it not unreasonable that this is a potential neutron star candidate. This source is also coincident with the bright emission associated with region 5 (see Figure \ref{regions}) and is consistent with both the powerlaw index derived from modelling both the emission from global and individual regions (Table \ref{fits}). Making it possible that the powerlaw component we observe arises from an extended PWN associated with source 8. However, deeper observations of the remnant using \textit{XMM-Newton} and \textit{NuSTAR} will allow us to disentangle this contribution from the thermal emission of the remnant.

Apart from a young thermally emitting neutron star, this soft X-ray point source could also potentially arise from clumps of ejecta. However due to the low number of counts, we are unable to detect emission line representative of ejecta emission. It is therefore difficult to differentiate between these two cases, and deeper observations of this source, as well as a timing analysis would be able to shed light on the nature of this source.

\section{Conclusion}
In this paper we present \emph{XMM-Newton} observations of
the MM SNR G346.6--0.2. We perform imaging and spectral analysis to
characterise the properties of the remnant. We find that the remnant shows
bright central emission fully enclosed by the radio shell, similar to that of
other MM SNRs. The X-ray emission is relatively clumpy in nature, with the bulk of the
soft X-rays found towards the
north of the remnant, while the emission overlapping the position of the OH
masers is quite hard in nature. As found in a previous \emph{Suzaku}
observation, we confirm that the X-ray spectrum of the SNR is well described by a
cold ($0.21-0.28$\,keV) recombining thermal plasma with sub-solar abundances of
Mg, Si and S. But unlike some previous studies we also find that all regions
(except for region 2) require either an additional power-law component with a photon
index of $\sim2$, or a thermal APEC component with a temperature of $\sim2.0$ keV.

We investigated the possible origin of this hard X-ray tail and find that it either arises from the GRXE, an unidenitified PWN, or synchrotron X-ray emission from a population of electrons accelerated by the shock front.

If this emission results from the GRXE, it is likely this hard X-ray tail arises from thermal CIE emission arising from faint coronal X-ray sources \citep{2012ApJ...753..129Y} or from the low temperature component of the GRXE \citep[see e.g.,][]{2013PASJ...65...19U}.

Based on its morphology, its Galactic latitude, the density of the surrounding environment and its association with a dense molecular cloud, G346.6--0.2 most likely arises from a massive progenitor that underwent CC. As such it is possible that this hard X-ray trail arise from an unidentified PWN. We performed a point source analysis of the sources within the field of view of the detector in an attempt to find a neutron star candidate. Defining a circular radius for which we would expect to find an associated neutron star given a reasonable kick velocity of a few 100 km s$^{-1}$, we find one source (Source 8) has no optical or IR counterparts, a photon index and HR comparable to that of young neutron stars with a PWN and a column density similar to the remnant. However, due to the relatively short exposure time of the current \textit{XMM} observation, deeper observations of the source would be needed confirm the origin of this point source.

Using the shock velocity we derived using the \citet{1991ApJ...373..543W} model and using the magnetic field value derived from Zeeman splitting measurements as an \textit{upperlimit} to the magnetic field found within the remnant, it is possible that this emission could also arise from synchrotron
X-ray emission from a population of electrons being accelerated by the shock
front. The unique properties of this source, in addition to the possible radiative nature of its shock front and the presence of a dense environment could possibly aid in the production of X-ray synchrotron emission from a remnant that one would not expect to observe this type of emission. If confirmed, this would make G346.6--0.2 an important new object in the class of synchrotron emitting SNRs. 

\acknowledgements
We thank the anonymous referee for their helpful comments and suggestions that improved the quality of the paper.  This work was based on observations obtained with \emph{XMM-Newton},
an ESA science mission with instruments and contributions directly funded by ESA Member States and NASA. \\
\textit{Facilities: XMM (EPIC)}. \textit{Software: XMMSAS}\\

\bibliography{g346bib}

\input{snr_fit.tex}

\input{pointsrc_radec_counterparts.tex}

\input{point_src_table.tex}

\end{document}

%% file: snr_fit.tex
\setlength{\tabcolsep}{0.02in}
\begin{deluxetable*}{cccccccccccc}
\tablewidth{0.95\textwidth}
  \tablecaption{Best-fit Parameters for the Global Spectrum and Individual Regions Using an Absorbed VRNEI+POWERLAW Model.\label{fits}}
\tablehead{
  \colhead{Region}&\colhead{$N_{\rm H}$ }& \colhead{$kT$}&\colhead{$kT_{\rm init}$}&\colhead{Mg}&\colhead{Si}&\colhead{S}&\colhead{$\tau$=$n_{e}t$}&\colhead{$F_{\rm vrnei}$\tablenotemark{b,c}}&\colhead{$\Gamma$}&\colhead{$F_{\rm pwl}$\tablenotemark{b,c} }&\colhead{$\chi^2_\nu\ (dof)$}\\
   &\colhead{($10^{22}$\,cm$^{-2}$)}&\colhead{(keV)}&\colhead{(keV)}&&&&\colhead{($10^{11}$\,cm$^{-3}$\,s)}&&&& 
}
\startdata
Global & $3.1_{-0.2}^{+0.1} $ & $0.26\pm0.02$ & $6^{+4}_{-1}$  & $0.38\pm0.1$ & $ 0.73\pm0.1$ & $0.60\pm0.01$ & $5.4\pm0.7$ & $11\pm2$ & $2.0^{+0.7}_{-0.9}$ & $3.3\pm2$ & 0.99 (922) \\
1 & $3.3\pm0.2$ & $0.23\pm0.02$ & 6\tablenotemark{a}  & $0.12\pm0.1$ & $ 0.46\pm0.07$ & $0.38^{+0.09}_{-0.07}$ & $5.3^{+0.8}_{-0.6}$ &$5.4^{+0.5}_{-1.0}$  &$1.8^{+1.0}_{-1.1}$ &$1.6^{+5}_{-2}$ &1.04 (651) \\
2 & $2.9\pm0.2$ & $0.24\pm0.02$ & $6\tablenotemark{a}$  & $0.17\pm0.1$ & $ 0.63^{+0.09}_{-0.08}$ & $0.50^{+0.08}_{-0.07}$ & $3.8\pm0.5$ &$9.4\pm0.03$  &\nodata & \nodata&0.99 (710) \\
3 & $2.0\pm0.3$ & $0.21\pm0.04$  & 6\tablenotemark{a} & \nodata  &
\nodata & \nodata & $4.0^{+0.8}_{-1.1}$ & $5.3\pm2$ & $1.0^{+1.1}_{-0.8}$ &$3.7^{+6}_{-5}$  & 1.01 (836) \\
4 & $3.4\pm0.2$ & $0.26\pm0.02$  & 6\tablenotemark{a} & $0.23\pm0.14$  & $0.56^{+0.09}_{-0.08}$ & $0.51^{+0.07}_{-0.09}$ & $5.3^{+0.7}_{-0.6}$& $12^{+2}_{-5}$  & $1.5^{+0.9}_{-1.2}$& $4.0^{+7}_{-2}$  & 1.03 (973) \\
5 & $2.8^{+0.2}_{-0.3}$ & $0.24\pm0.04$  & 6\tablenotemark{a} &
$0.29\pm0.2$  & \nodata & $0.54^{+0.2}_{-0.1}$ & $4.3^{+0.9}_{-0.4}$ & $5.7^{+2}_{-1}$ & $2.5^{+0.4}_{-1.2}$ &$3.5^{+3}_{-2}$& 1.04 (585)  \\
6 & $2.9^{+0.2}_{-0.3}$ & $0.28\pm0.03$  & 6\tablenotemark{a} &
$0.42\pm0.2$  & \nodata & \nodata & $5.4\pm0.8$ &
$5.8^{+2}_{-1}$ & $2.1^{+0.4}_{-0.8}$ & $4.1\pm3$ & 1.03 (621) 
\enddata
\tablecomments{All uncertainties correspond to 90\% confidence level.}
\tablenotetext{a}{We fix $kT_{\rm init}$ for all individual regions at the
global best-fit value.}
\tablenotetext{b}{Absorbed X-ray fluxes in the 0.5--7\,keV energy range.}
\tablenotetext{c}{Flux units: $10^{-13}$ erg\,cm$^{-2}$\,s$^{-1}$}
\end{deluxetable*}

\setlength{\tabcolsep}{0.02in}
\begin{deluxetable*}{cccccccccccc}
	\tablewidth{0.95\textwidth}
	\tablecaption{Best-fit Parameters for the Global Spectrum and Individual Regions Using an Absorbed VRNEI+APEC Model.\label{fitsapec}}
	\tablehead{
		\colhead{Region}&\colhead{$N_{\rm H}$ }& \colhead{$kT$}&\colhead{$kT_{\rm init}$}&\colhead{Mg}&\colhead{Si}&\colhead{S}&\colhead{$\tau$=$n_{e}t$}&\colhead{$F_{\rm vrnei}$\tablenotemark{b,c}}&\colhead{$kT_{\rm apec}$}&\colhead{$F_{\rm apec}$\tablenotemark{b,c} }&\colhead{$\chi^2_\nu\ (dof)$}\\
		&\colhead{($10^{22}$\,cm$^{-2}$)}&\colhead{(keV)}&\colhead{(keV)}&&&&\colhead{($10^{11}$\,cm$^{-3}$\,s)}&&&& 
	}
	\startdata
	Global & $3.0\pm0.2$ & $0.24\pm0.02$ & $6^{+9}_{-2}$  & $0.34\pm0.1$ & $ 0.68\pm0.1$ & $0.49\pm0.1$ &$5.3\pm0.6$ & $8.8\pm0.05$ & \textbf{$1.9^{+0.6}_{-0.3}$} & $5.2\pm0.3$ & 0.96 (899) \\
		1 & $3.3\pm0.2$&$0.23\pm0.02$&$6\tablenotemark{a}$ &$0.08_{-0.04}^{+0.10}$ &$0.45\pm0.1$&$0.36\pm0.1$& $5.8\pm1$& $5.0^{+0.2}_{-0.6}$&$2.4\pm0.4$&$2.3\pm0.2$& 1.06 (651)\\
			2 & $2.9\pm0.2$ & $0.24\pm0.02$ & $6\tablenotemark{a}$  & $0.17\pm0.1$ & $ 0.63^{+0.09}_{-0.08}$ & $0.50^{+0.08}_{-0.07}$ & $3.8\pm0.5$ &$9.4\pm0.03$  &\nodata & \nodata&0.99 (710) \\
			3 & $2.00\pm0.2$&$0.21\pm0.03$&$6\tablenotemark{a}$ &\nodata&\nodata&\nodata&$4.0^{+0.8}_{-1.1}$&$4.5^{+0.8}_{-0.2}$&$2.5_{-0.7}^{+10}$&$3.6\pm0.3$&1.03 (826)\\
			4 & $3.5\pm0.2$ & $0.22\pm0.2$& $6\tablenotemark{a}$ & $0.22\pm0.2$ &$0.54\pm0.1$ & $0.50\pm0.1$ & $6.0^{+0.7}_{-0.6}$&$8.8^{+0.3}_{-0.5}$&$1.7\pm0.2$&$5.2\pm0.2$&1.02 (969)\\
	5 & $2.7\pm0.2$&$0.24\pm0.03$&$6\tablenotemark{a}$ &$0.28\pm0.2$ &\nodata&$0.57\pm0.1$& $4.3\pm0.6$& $5.6^{+0.3}_{-0.7}$& $2.3^{-0.5}_{+6.0}$&$2.8\pm0.1$& 1.04 (585)\\
	6&$2.9^{+0.2}_{-0.3}$&$0.28\pm0.03$&$6\tablenotemark{a}$ &$0.43\pm0.2$&\nodata&\nodata&$5.8\pm0.9$&$5.3^{-0.8}_{+1.4}$&$3.4_{-1.1}^{+6.0}$&$3.8\pm0.1$&1.03 (621)
	\enddata
	\tablecomments{All uncertainties correspond to 90\% confidence level.}
	\tablenotetext{a}{We fix $kT_{\rm init}$ for all individual regions at the
		global best-fit value.}
	\tablenotetext{b}{Absorbed X-ray fluxes in the 0.5--7\,keV energy range.}
	\tablenotetext{c}{Flux units: $10^{-13}$ erg\,cm$^{-2}$\,s$^{-1}$}
\end{deluxetable*}

%% file: pointsrc_radec_counterparts.tex
\clearpage
\begin{deluxetable*}{ccccccc}
  \tablecaption{Position of the optical/UV/X-ray counterparts found within 3$\sigma$ of the X-ray sources in the field of view.\label{counterparts}}
\tabletypesize{\small}
  \tablewidth{0pt} 
\tablehead{
\colhead{Src} & \colhead{R.A.}& \colhead{Dec.}  & \colhead{Positional} &\colhead{UNSO-B1\tablenotemark{a}} & \colhead{2MASS\tablenotemark{a}} & \colhead{3XMM}\\
 & & & \colhead{Uncert. (\arcsec)} & & &
 }\\
\startdata
1& 17:10:52.32&$-$40:02:35.52 &0.5& \nodata &\nodata  & \nodata \\
2& 17:10:37.44&$-$40:05:31.56 &0.5& \nodata & \nodata & J171037.4$-$400531 \\
3& 17:10:17.76&$-$40:06:12.24 &0.4& \nodata & \nodata &  \nodata\\
4& 17:10:00.24&$-$40:07:13.08 &0.8& 0498$-$0501194 & 17100036$-$4007137 & J171000.4$-$400714 \\
5& 17:10:18.00&$-$40:08:20.40 &3.0& 0498$-$0501600 & 17101812$-$4008233 &\nodata  \\
6& 17:10:22.80&$-$40:08:47.40 &0.5& 0498$-$0501699 & 17102280$-$4008484 & J171022.8$-$400847 \\
7& 17:10:11.52&$-$40:09:55.08 &1.0& \nodata &\nodata  & \nodata \\
8& 17:10:14.88&$-$40:11:31.20 &1.5& \nodata & \nodata & \nodata \\
9& 17:09:43.44&$-$40:13:18.12 &0.6& 0497$-$0499796 & 17094361$-$4013188 & J170943.6$-$401318 \\
10& 17:09:33.84&$-$40:17:40.92 &0.6& 0497$-$0499624 & 17093387$-$4017406 & J170933.8$-$401740 \\
11& 17:10:08.64&$-$40:20:27.60 &1.5& \nodata & \nodata & J171008.7$-$402027 \\
12& 17:11:07.68&$-$40:14:13.20 &1.8& \nodata & \nodata & J171107.9$-$401414 \\
13& 17:10:56.88&$-$40:12:16.92 &1.0& 0497$-$0501494 & 17105706$-$4012172 & J171057.0$-$401217 \\
14& 17:11:01.20&$-$40:14:57.84 &1.0& \nodata & \nodata & J171101.2$-$401457 \\
15& 17:10:56.88&$-$40:20:03.84 &1.2& \nodata & \nodata & J171057.0$-$402002 \\
16& 17:10:10.56&$-$40:22:21.00 &5.1&\nodata  & 17101009$-$4022164& J171010.7$-$402210 \\
17& 17:09:49.92&$-$40:20:22.20 &4.1& \nodata & 17094990$-$4020205& \nodata \\
18& 17:10:33.36&$-$40:13:08.40 &4.0& 0497$-$0500904 & 17103301$-$4013134& J171033.4$-$401318 \\
19& 17:10:13.20&$-$40:13:49.08 &1.8&\nodata  & 17101317$-$4013527 & \nodata \\
20& 17:10:23.76&$-$40:12:19.08 &3.3& 0497$-$0500702 & 17102375$-$4012188 & \nodata \\
21& 17:10:29.28&$-$40:03:51.12 &1.2& 0499$-$0506406 & 17102939$-$4003507 & J171029.4$-$400350 \\
22& 17:10:21.60&$-$39:59:49.20 &1.2& \nodata & \nodata & J171021.6$-$395950 \\
23& 17:09:42.48&$-$39:58:44.04 &1.0& 0500$-$0505315 & 17094245$-$3958433 & J170942.4$-$395845 \\
24& 17:10:22.56&$-$40:07:38.28 &4.0& 0498$-$0501684 & 17102235$-$4007330 & J171022.7$-$400733\\
25&17:10:39.360&$-$40:12:16.92&0.9&\nodata&\nodata&J171039.4$-$401216\\
\enddata
\end{deluxetable*}
\clearpage

%% file: point_src_table.tex
\clearpage
\begin{deluxetable*}{ccccccccccccc}
  \tablecaption{\textbf{Properties of the X-ray Point Sources in the Field of View Detected with a Likelihood Threshold of 30$\sigma$.}\label{ptsrctable}}
  \tablewidth{0pt} 
\tablehead{
\colhead{Src} & 
\colhead{$R_{0.2-2}$\tablenotemark{a}} &
\colhead{$R_{2-10}$\tablenotemark{b}} & \colhead{Hardness} &
\colhead{$N_{\rm H}$} & \colhead{$\Gamma$} &
\colhead{$\chi^2_\nu$ (dof)} &
\colhead{$F_X$\tablenotemark{d} }   \\
&  
\colhead{($10^{-3}$\,cnt\,s$^{-1}$)} &
\colhead{($10^{-3}$\,cnt\,s$^{-1}$)} &
\colhead{Ratio\tablenotemark{c}} & \colhead{$(10^{22}$\,cm$^{-2}$)} & & &
\colhead{($10^{-14}$ erg\,cm$^{-2}$\,s$^{-1}$)}}

\startdata
1&$3.0\pm0.4$&$3.7\pm0.5$&0.1& $ 1.3 _{ -0.6 }^{+ 0.9 }$ & $ 2.2 _{ -0.6 }^{+ 0.8 } $ & $ 0.8\, (14) $ & $ 5.4 _{- 1}^{+ 0.4}$ \\ 
2&$4.3\pm0.5$&$1.7\pm0.3$&0.5& $ 0.5_{ -0.4}^{+ 2}$ & $ 0.6_{ -0.4 }^{+ 0.7} $ & $ 1.2 \,(14) $ & $ 3.4 _{- 0.9}^{+ 0.2} $ \\
3&$0.9\pm0.3$&$7.4\pm0.6$&0.8& $ 1.6 _{ -2}^{+ 3}$ & $ 0.2 _{ -0.5 }^{+ 0.6 } $ & $ 1.1\, (21) $ & $ 3.5 \pm 0.3 $ \\ 
4&$3.3\pm0.4$&$0.5\pm0.3$& $-0.7$ & $ 0.4 _{ -0.4 }^{+ 0.6 } $ & $3.1 _{ -1}^{+ 2}$ & $ 1.0\, (9) $ & $ 1.3 _{- 0.1}^{+ 0.1}$ \\
5&$2.2\pm0.4$&$0.8\pm0.3$& $-0.5$ & $ 4.0 _{ -2}^{+ 4 }  $ & $ 6.3 \pm 3 $ & $ 2.5\, (8) $ & $ 0.9 \pm0.5$ \\
6&$7.2\pm0.6$&$0.6\pm0.3$& $-0.9$ & 0.01$^{e}$ & $ 2.5 _{ -0.3 }^{+ 0.4 } $ & $ 1.0 \,(19) $ & $ 2.5 \pm 0.2$ \\
7&$3.1\pm0.4$&$0.7\pm0.3$& $-0.6$ & $ 2.3 \pm2 $ & $ 4.9 _{ -3}^{+ 4}$ & $ 1.9 \,(8) $ & $ 1.4 \pm0.7 $ \\
8&$3.7\pm0.4$&$2.4\pm0.4$& $-0.2$ & $ 3.4 _{ -1}^{+ 2 }   $ & $ 4.4 _{ -2}^{+ 1} $ & $ 0.8 \,(15) $ & $ 2.7 \pm0.3$ \\
9&$4.8\pm0.5$&$0.0\pm0.2$& $-1.0$ & $ 1.0 _{ -0.7 }^{+ 0.5 } $ & $ 8.0 \pm 3 $ & $ 0.9 \,(10) $ & $ 1.8 \pm0.4$ \\
10&$3.1\pm0.4$&$0.0\pm0.2$& $-1.0 $& $ 0.3 _{ -0.3 }^{+ 1}$ & $ 5.5 \pm 2$ & $ 1.34 \,(6) $ & $ 2.7 \pm1$ \\
11&$0.0\pm0.2$&$0.0\pm0.2$& $0.0$ &\nodata &\nodata & \nodata& \nodata \\
12&$0.4\pm0.2$&$1.5\pm0.3$&0.6& $ 1.8 _{ -1.8}^{+ 7}$ & $ 1.2 _{ -1}^{+ 2}$ & $ 0.8\, (6) $ & $ 1.8 \pm0.6$ \\
13&$2.0\pm0.3$&$0.4\pm0.3$& $-0.7$ & $ 1.9 _{ -1.8}^{+ 3}$ & $ 5.0 _{ -2}^{+ 3}$ & $ 1.4 \,(5) $ & $ 0.9 \pm0.2$ \\
14&$0\pm0.2$&$1.9\pm0.4$&1.0& $ 3.2 _{ -3}^{+ 9}$ & $ 0.7 _{ -1}^{+ 2}$ & $ 0.4 \,(6) $ & $ 1.8\pm0.5$ \\
15&$0.3\pm0.2$&$0.9\pm0.3$&0.6& $ 2.1 _{ -2}^{+ 12 } $ & $ 1.8 _{ -1}^{+ 4}$ & $ 0.2\, (4) $ & $ 2.2 \pm0.8$ \\
16&$0.0\pm0.1$&$0.2\pm0.3$&1.0&\nodata &\nodata & \nodata & \nodata \\
17&$0.0\pm0.2$&$0.2\pm0.3$&1.0&\nodata &\nodata & \nodata & \nodata \\
18&$0.8\pm0.3$&$1.5\pm0.3$&0.3& $ 0.6 _{ -0.6}^{+ 4}$ & $ 1.9_{ -0.8}^{+ 2 } $ & $ 1.5\, (6) $ & $ 1.1 \pm0.4$ \\
19&$1.4\pm0.3$&$1.0\pm0.3$& $-0.2$& $ 3.3 _{ -2}^{+ 5}$ & $ 3.8 _{ -2}^{+ 3}$ & $ 1.0\, (6) $ & $ 1.3 \pm0.3$ \\
20&$2.5\pm0.4$&$1.6\pm0.4$& $-0.2$ & $ 0.8 _{ -0.7}^{+ 1}$ & $ 1.8 _{ -0.7}^{+ 0.9}$ & $ 1.5\, (11) $ & $ 2.1\pm0.3$ \\
21&$1.2\pm0.3$&$0.0\pm0.2$& $-1.0$& $ 1.5_{ -0.9 }^{+ 0.8}$ & $ 10 _{ -6}^{+ 1}$ & $ 1.5\, (3) $ & $ 0.4 \pm1.1$ \\
22&$0.7\pm0.3$&$0.5\pm0.3$& $-0.2$ & $ 0.01^{e}$ & $ 1.4 _{ -1}^{+ 2}$ & $ 0.5\, (4) $ & $ 1.6 \pm0.1$ \\
23&$1.2\pm0.3$&$0.4\pm0.3$&$ -0.5$ & $  1.6_{ -0.6 }^{+ 0.5}$ & $ 10 _{ -6}^{+ 1}$ & $ 1.6\, (2) $ & $ 0.9 \pm 0.2$ \\
24&$2.0\pm0.3$&$1.1\pm0.3$& $-0.3$ & $ 2.5 _{ -2}^{+ 3}$ & $ 3.6 _{ -2}^{+ 3}$ & $ 0.3\,(6) $ & $ 1.6 \pm0.5$ \\
25&$0.4\pm0.3$&$3.5\pm0.5$& 0.80 & $ 20 _{ -18}^{+ 15}$ & $ 3.5 _{ -3}^{+ 4}$ & $ 1.3\,(9) $ & $ 2.0 \pm0.7$ \\
\enddata
\tablecomments{A TBABS*POWERLAW model was used to model the X-ray emission from these sources. All uncertainties are at the 90\% confidence level. The lower uncertainty for a number of these sources were unconstrained due to the low number of counts, thus the lower uncertainty for their fluxes is zero.}
\tablenotetext{a}{Count rate in 0.2--2\,keV band derived from filtered PN data.}
\tablenotetext{b}{Count rate in 2--10\,keV band derived from filtered PN data.}
\tablenotetext{c}{Hardness ratio defined as HR$=(R_{2-10}-R_{0.2-2})/(R_{2-10}+R_{0.2-2})$, where $R$ is the count rate. Sources with HR $< 0$ and $>0$ are classified as soft and hard X-ray sources, respectively.}
\tablenotetext{d}{Absorbed flux in the 0.5--5.0\,keV energy band.}
\tablenotetext{e}{$N_{\rm H}$ for these sources are $\sim 10^{20}$\,cm$^{-2}$. We therefore fixed $N_{\rm H}$ at $0.01\times10^{22}$\,cm$^{-2}$}
\end{deluxetable*}
\clearpage

%% file: g346_ms.bbl
\begin{thebibliography}{}
\expandafter\ifx\csname natexlab\endcsname\relax\def\natexlab#1{#1}\fi

\bibitem[{{Acero} {et~al.}(2013){Acero}, {Gallant}, {Ballet}, {Renaud}, \&
  {Terrier}}]{2013A&A...551A...7A}
{Acero}, F., {Gallant}, Y., {Ballet}, J., {Renaud}, M., \& {Terrier}, R. 2013,
  \aap, 551, A7

\bibitem[{{Aharonian} \& {Atoyan}(1999)}]{1999A&A...351..330A}
{Aharonian}, F.~A., \& {Atoyan}, A.~M. 1999, \aap, 351, 330

\bibitem[{{Anders} \& {Grevesse}(1989)}]{1989GeCoA..53..197A}
{Anders}, E., \& {Grevesse}, N. 1989, \gca, 53, 197

\bibitem[{{Andersen} {et~al.}(2011){Andersen}, {Rho}, {Reach}, {Hewitt}, \&
  {Bernard}}]{2011ApJ...742....7A}
{Andersen}, M., {Rho}, J., {Reach}, W.~T., {Hewitt}, J.~W., \& {Bernard}, J.~P.
  2011, \apj, 742, 7

\bibitem[{{Anderson} {et~al.}(2014){Anderson}, {Gaensler}, {Kaplan}, {Slane},
  {Muno}, {Posselt}, {Hong}, {Murray}, {Steeghs}, {Brogan}, {Drake}, {Farrell},
  {Benjamin}, {Chakrabarty}, {Drew}, {Finley}, {Grindlay}, {Lazio}, {Lee},
  {Mauerhan}, \& {van Kerkwijk}}]{2014ApJS..212...13A}
{Anderson}, G.~E., {Gaensler}, B.~M., {Kaplan}, D.~L., {et~al.} 2014, \apjs,
  212, 13

\bibitem[{{Arzoumanian} {et~al.}(2002){Arzoumanian}, {Chernoff}, \&
  {Cordes}}]{2002ApJ...568..289A}
{Arzoumanian}, Z., {Chernoff}, D.~F., \& {Cordes}, J.~M. 2002, \apj, 568, 289

\bibitem[{{Aschenbach}(1998)}]{1998Natur.396..141A}
{Aschenbach}, B. 1998, \nat, 396, 141

\bibitem[{{Auchettl} {et~al.}(2015){Auchettl}, {Slane}, {Castro}, {Foster}, \&
  {Smith}}]{2015ApJ...810...43A}
{Auchettl}, K., {Slane}, P., {Castro}, D., {Foster}, A.~R., \& {Smith}, R.~K.
  2015, \apj, 810, 43

\bibitem[{{Ballet}(2006)}]{2006AdSpR..37.1902B}
{Ballet}, J. 2006, Advances in Space Research, 37, 1902

\bibitem[{{Bamba} {et~al.}(2000){Bamba}, {Koyama}, \&
  {Tomida}}]{2000PASJ...52.1157B}
{Bamba}, A., {Koyama}, K., \& {Tomida}, H. 2000, \pasj, 52, 1157

\bibitem[{{Bocchino} \& {Bykov}(2001)}]{2001A&A...376..248B}
{Bocchino}, F., \& {Bykov}, A.~M. 2001, \aap, 376, 248

\bibitem[{{Borkowski} {et~al.}(2010){Borkowski}, {Reynolds}, {Green}, {Hwang},
  {Petre}, {Krishnamurthy}, \& {Willett}}]{2010ApJ...724L.161B}
{Borkowski}, K.~J., {Reynolds}, S.~P., {Green}, D.~A., {et~al.} 2010, \apjl,
  724, L161

\bibitem[{{Borkowski} {et~al.}(2001){Borkowski}, {Rho}, {Reynolds}, \&
  {Dyer}}]{2001ApJ...550..334B}
{Borkowski}, K.~J., {Rho}, J., {Reynolds}, S.~P., \& {Dyer}, K.~K. 2001, \apj,
  550, 334

\bibitem[{{Broersen} {et~al.}(2014){Broersen}, {Chiotellis}, {Vink}, \&
  {Bamba}}]{2014MNRAS.441.3040B}
{Broersen}, S., {Chiotellis}, A., {Vink}, J., \& {Bamba}, A. 2014, \mnras, 441,
  3040

\bibitem[{{Brogan} {et~al.}(2000){Brogan}, {Frail}, {Goss}, \&
  {Troland}}]{2000ApJ...537..875B}
{Brogan}, C.~L., {Frail}, D.~A., {Goss}, W.~M., \& {Troland}, T.~H. 2000, \apj,
  537, 875

\bibitem[{{Brogan} {et~al.}(2013){Brogan}, {Goss}, {Hunter}, {Richards},
  {Chandler}, {Lazendic}, {Koo}, {Hoffman}, \&
  {Claussen}}]{2013ApJ...771...91B}
{Brogan}, C.~L., {Goss}, W.~M., {Hunter}, T.~R., {et~al.} 2013, \apj, 771, 91

\bibitem[{{Butt} {et~al.}(2001){Butt}, {Torres}, {Combi}, {Dame}, \&
  {Romero}}]{2001ApJ...562L.167B}
{Butt}, Y.~M., {Torres}, D.~F., {Combi}, J.~A., {Dame}, T., \& {Romero}, G.~E.
  2001, \apjl, 562, L167

\bibitem[{{Bykov}(2002)}]{2002cosp...34E.970B}
{Bykov}, A. 2002, 34th COSPAR Scientific Assembly, 34

\bibitem[{{Cassam-Chena{\"i}} {et~al.}(2004{\natexlab{a}}){Cassam-Chena{\"i}},
  {Decourchelle}, {Ballet}, {Hwang}, {Hughes}, {Petre}, \& {et
  al.}}]{2004A&A...414..545C}
{Cassam-Chena{\"i}}, G., {Decourchelle}, A., {Ballet}, J., {et~al.}
  2004{\natexlab{a}}, \aap, 414, 545

\bibitem[{{Cassam-Chena{\"i}} {et~al.}(2004{\natexlab{b}}){Cassam-Chena{\"i}},
  {Decourchelle}, {Ballet}, {Sauvageot}, {Dubner}, \&
  {Giacani}}]{2004A&A...427..199C}
---. 2004{\natexlab{b}}, \aap, 427, 199

\bibitem[{{Chen} {et~al.}(2004){Chen}, {Su}, {Slane}, \&
  {Wang}}]{2004ApJ...616..885C}
{Chen}, Y., {Su}, Y., {Slane}, P.~O., \& {Wang}, Q.~D. 2004, \apj, 616, 885

\bibitem[{{Chevalier}(1977)}]{1977ARA&A..15..175C}
{Chevalier}, R.~A. 1977, \araa, 15, 175

\bibitem[{{Clark} {et~al.}(1975){Clark}, {Green}, \&
  {Caswell}}]{1975AuJPA.......75C}
{Clark}, D.~H., {Green}, A.~J., \& {Caswell}, J.~L. 1975, AJPAS, 75

\bibitem[{{Claussen} {et~al.}(1997){Claussen}, {Frail}, {Goss}, \&
  {Gaume}}]{1997ApJ...489..143C}
{Claussen}, M.~J., {Frail}, D.~A., {Goss}, W.~M., \& {Gaume}, R.~A. 1997, \apj,
  489, 143

\bibitem[{{Claussen} {et~al.}(1999){Claussen}, {Goss}, {Frail}, \&
  {Seta}}]{1999AJ....117.1387C}
{Claussen}, M.~J., {Goss}, W.~M., {Frail}, D.~A., \& {Seta}, M. 1999, \aj, 117,
  1387

\bibitem[{{Cox} {et~al.}(1999){Cox}, {Shelton}, {Maciejewski}, {Smith},
  {Plewa}, {Pawl}, \& {R{\'o}{\.z}yczka}}]{1999ApJ...524..179C}
{Cox}, D.~P., {Shelton}, R.~L., {Maciejewski}, W., {et~al.} 1999, \apj, 524,
  179

\bibitem[{{Cui} \& {Cox}(1992)}]{1992ApJ...401..206C}
{Cui}, W., \& {Cox}, D.~P. 1992, \apj, 401, 206

\bibitem[{{Dubner} {et~al.}(1993){Dubner}, {Moffett}, {Goss}, \&
  {Winkler}}]{1993AJ....105.2251D}
{Dubner}, G.~M., {Moffett}, D.~A., {Goss}, W.~M., \& {Winkler}, P.~F. 1993,
  \aj, 105, 2251

\bibitem[{{Dwarkadas}(2005)}]{2005ApJ...630..892D}
{Dwarkadas}, V.~V. 2005, \apj, 630, 892

\bibitem[{{Ebisawa} {et~al.}(2005){Ebisawa}, {Tsujimoto}, {Paizis},
  {Hamaguchi}, {Bamba}, {Cutri}, {Kaneda}, {Maeda}, {Sato}, {Senda}, {Ueno},
  {Yamauchi}, {Beckmann}, {Courvoisier}, {Dubath}, \&
  {Nishihara}}]{2005ApJ...635..214E}
{Ebisawa}, K., {Tsujimoto}, M., {Paizis}, A., {et~al.} 2005, \apj, 635, 214

\bibitem[{{Ergin} \& {Ercan}(2012)}]{2012AIPC.1505..265E}
{Ergin}, T., \& {Ercan}, E.~N. 2012, American Institute of Physics Conference
  Series, 1505, 265

\bibitem[{{Eriksen} {et~al.}(2011){Eriksen}, {Hughes}, {Badenes}, {Fesen},
  {Ghavamian}, {Moffett}, {Plucinksy}, {Rakowski}, {Reynoso}, \&
  {Slane}}]{2011ApJ...728L..28E}
{Eriksen}, K.~A., {Hughes}, J.~P., {Badenes}, C., {et~al.} 2011, \apjl, 728,
  L28

\bibitem[{{Foster} {et~al.}(2012){Foster}, {Ji}, {Smith}, \&
  {Brickhouse}}]{2012ApJ...756..128F}
{Foster}, A.~R., {Ji}, L., {Smith}, R.~K., \& {Brickhouse}, N.~S. 2012, \apj,
  756, 128

\bibitem[{{Frail} {et~al.}(1996){Frail}, {Giacani}, {Goss}, \&
  {Dubner}}]{1996ApJ...464L.165F}
{Frail}, D.~A., {Giacani}, E.~B., {Goss}, W.~M., \& {Dubner}, G. 1996, \apjl,
  464, L165

\bibitem[{{Gaensler} \& {Slane}(2006)}]{2006ARA&A..44...17G}
{Gaensler}, B.~M., \& {Slane}, P.~O. 2006, \araa, 44, 17

\bibitem[{{Gaensler} {et~al.}(2001){Gaensler}, {Slane}, {Gotthelf}, \&
  {Vasisht}}]{2001ApJ...559..963G}
{Gaensler}, B.~M., {Slane}, P.~O., {Gotthelf}, E.~V., \& {Vasisht}, G. 2001,
  \apj, 559, 963

\bibitem[{{Gelfand} {et~al.}(2013){Gelfand}, {Castro}, {Slane}, {Temim},
  {Hughes}, \& {Rakowski}}]{2013ApJ...777..148G}
{Gelfand}, J.~D., {Castro}, D., {Slane}, P.~O., {et~al.} 2013, \apj, 777, 148

\bibitem[{{Gotthelf} {et~al.}(2001){Gotthelf}, {Koralesky}, {Rudnick}, {Jones},
  {Hwang}, \& {Petre}}]{2001ApJ...552L..39G}
{Gotthelf}, E.~V., {Koralesky}, B., {Rudnick}, L., {et~al.} 2001, \apjl, 552,
  L39

\bibitem[{{Harrus} {et~al.}(1997){Harrus}, {Hughes}, {Singh}, {Koyama}, \&
  {Asaoka}}]{1997ApJ...488..781H}
{Harrus}, I.~M., {Hughes}, J.~P., {Singh}, K.~P., {Koyama}, K., \& {Asaoka}, I.
  1997, \apj, 488, 781

\bibitem[{{Hewitt} {et~al.}(2009){Hewitt}, {Rho}, {Andersen}, \&
  {Reach}}]{2009ApJ...694.1266H}
{Hewitt}, J.~W., {Rho}, J., {Andersen}, M., \& {Reach}, W.~T. 2009, \apj, 694,
  1266

\bibitem[{{Hobbs} {et~al.}(2005){Hobbs}, {Lorimer}, {Lyne}, \&
  {Kramer}}]{2005MNRAS.360..974H}
{Hobbs}, G., {Lorimer}, D.~R., {Lyne}, A.~G., \& {Kramer}, M. 2005, \mnras,
  360, 974

\bibitem[{{Huang} \& {Thaddeus}(1985)}]{1985ApJ...295L..13H}
{Huang}, Y.-L., \& {Thaddeus}, P. 1985, \apjl, 295, L13

\bibitem[{{Hwang} {et~al.}(2002){Hwang}, {Decourchelle}, {Holt}, \&
  {Petre}}]{2002ApJ...581.1101H}
{Hwang}, U., {Decourchelle}, A., {Holt}, S.~S., \& {Petre}, R. 2002, \apj, 581,
  1101

\bibitem[{{Itoh}(1977)}]{1977PASJ...29..813I}
{Itoh}, H. 1977, \pasj, 29, 813

\bibitem[{{Itoh} \& {Masai}(1989)}]{1989MNRAS.236..885I}
{Itoh}, H., \& {Masai}, K. 1989, \mnras, 236, 885

\bibitem[{{Iwamoto} {et~al.}(1999){Iwamoto}, {Brachwitz}, {Nomoto},
  {Kishimoto}, {Umeda}, {Hix}, \& {Thielemann}}]{1999ApJS..125..439I}
{Iwamoto}, K., {Brachwitz}, F., {Nomoto}, K., {et~al.} 1999, \apjs, 125, 439

\bibitem[{{Kaneda} {et~al.}(1997){Kaneda}, {Makishima}, {Yamauchi}, {Koyama},
  {Matsuzaki}, \& {Yamasaki}}]{1997ApJ...491..638K}
{Kaneda}, H., {Makishima}, K., {Yamauchi}, S., {et~al.} 1997, \apj, 491, 638

\bibitem[{{Kargaltsev} \& {Pavlov}(2008)}]{2008AIPC..983..171K}
{Kargaltsev}, O., \& {Pavlov}, G.~G. 2008, in American Institute of Physics
  Conference Series, Vol. 983, 40 Years of Pulsars: Millisecond Pulsars,
  Magnetars and More, ed. C.~{Bassa}, Z.~{Wang}, A.~{Cumming}, \& V.~M.
  {Kaspi}, 171--185

\bibitem[{{Katsuda} {et~al.}(2009){Katsuda}, {Petre}, {Hwang}, {Yamaguchi},
  {Mori}, \& {Tsunemi}}]{2009PASJ...61S.155K}
{Katsuda}, S., {Petre}, R., {Hwang}, U., {et~al.} 2009, \pasj, 61, S155

\bibitem[{{Katsuda} {et~al.}(2015){Katsuda}, {Acero}, {Tominaga}, {Fukui},
  {Hiraga}, {Koyama}, {Lee}, {Mori}, {Nagataki}, {Ohira}, {Petre}, {Sano},
  {Takeuchi}, {Tamagawa}, {Tsuji}, {Tsunemi}, \&
  {Uchiyama}}]{2015ApJ...814...29K}
{Katsuda}, S., {Acero}, F., {Tominaga}, N., {et~al.} 2015, \apj, 814, 29

\bibitem[{{Kawasaki} {et~al.}(2005){Kawasaki}, {Ozaki}, {Nagase}, {Inoue}, \&
  {Petre}}]{2005ApJ...631..935K}
{Kawasaki}, M., {Ozaki}, M., {Nagase}, F., {Inoue}, H., \& {Petre}, R. 2005,
  \apj, 631, 935

\bibitem[{{Kawasaki} {et~al.}(2002){Kawasaki}, {Ozaki}, {Nagase}, {Masai},
  {Ishida}, \& {Petre}}]{2002ApJ...572..897K}
{Kawasaki}, M.~T., {Ozaki}, M., {Nagase}, F., {et~al.} 2002, \apj, 572, 897

\bibitem[{{Keohane} {et~al.}(2007){Keohane}, {Reach}, {Rho}, \&
  {Jarrett}}]{2007ApJ...654..938K}
{Keohane}, J.~W., {Reach}, W.~T., {Rho}, J., \& {Jarrett}, T.~H. 2007, \apj,
  654, 938

\bibitem[{{Koralesky} {et~al.}(1998){Koralesky}, {Frail}, {Goss}, {Claussen},
  \& {Green}}]{1998AJ....116.1323K}
{Koralesky}, B., {Frail}, D.~A., {Goss}, W.~M., {Claussen}, M.~J., \& {Green},
  A.~J. 1998, \aj, 116, 1323

\bibitem[{{Koyama} {et~al.}(1997){Koyama}, {Kinugasa}, {Matsuzaki},
  {Nishiuchi}, {Sugizaki}, {Torii}, {Yamauchi}, \&
  {Aschenbach}}]{1997PASJ...49L...7K}
{Koyama}, K., {Kinugasa}, K., {Matsuzaki}, K., {et~al.} 1997, \pasj, 49, L7

\bibitem[{{Koyama} {et~al.}(1995){Koyama}, {Petre}, {Gotthelf}, {Hwang},
  {Matsuura}, {Ozaki}, \& {Holt}}]{1995Natur.378..255K}
{Koyama}, K., {Petre}, R., {Gotthelf}, E.~V., {et~al.} 1995, \nat, 378, 255

\bibitem[{{Lazendic} \& {Slane}(2006)}]{2006ApJ...647..350L}
{Lazendic}, J.~S., \& {Slane}, P.~O. 2006, \apj, 647, 350

\bibitem[{{Lazendic} {et~al.}(2004){Lazendic}, {Slane}, {Gaensler}, {Reynolds},
  {Plucinsky}, \& {Hughes}}]{2004ApJ...602..271L}
{Lazendic}, J.~S., {Slane}, P.~O., {Gaensler}, B.~M., {et~al.} 2004, \apj, 602,
  271

\bibitem[{{Long} {et~al.}(1991){Long}, {Blair}, {Matsui}, \&
  {White}}]{1991ApJ...373..567L}
{Long}, K.~S., {Blair}, W.~P., {Matsui}, Y., \& {White}, R.~L. 1991, \apj, 373,
  567

\bibitem[{{Lopez} {et~al.}(2013){Lopez}, {Pearson}, {Ramirez-Ruiz}, {Castro},
  {Yamaguchi}, {Slane}, \& {Smith}}]{2013ApJ...777..145L}
{Lopez}, L.~A., {Pearson}, S., {Ramirez-Ruiz}, E., {et~al.} 2013, \apj, 777,
  145

\bibitem[{{Lopez} {et~al.}(2009){Lopez}, {Ramirez-Ruiz}, {Badenes},
  {Huppenkothen}, {Jeltema}, \& {Pooley}}]{2009ApJ...706L.106L}
{Lopez}, L.~A., {Ramirez-Ruiz}, E., {Badenes}, C., {et~al.} 2009, \apjl, 706,
  L106

\bibitem[{{Lopez} {et~al.}(2011){Lopez}, {Ramirez-Ruiz}, {Huppenkothen},
  {Badenes}, \& {Pooley}}]{2011ApJ...732..114L}
{Lopez}, L.~A., {Ramirez-Ruiz}, E., {Huppenkothen}, D., {Badenes}, C., \&
  {Pooley}, D.~A. 2011, \apj, 732, 114

\bibitem[{{Lopez} {et~al.}(2015){Lopez}, {Grefenstette}, {Reynolds}, {An},
  {Boggs}, {Christensen}, {Craig}, {Eriksen}, {Fryer}, {Hailey}, {Harrison},
  {Madsen}, {Stern}, {Zhang}, \& {Zoglauer}}]{2015ApJ...814..132L}
{Lopez}, L.~A., {Grefenstette}, B.~W., {Reynolds}, S.~P., {et~al.} 2015, \apj,
  814, 132

\bibitem[{{Lovchinsky} {et~al.}(2011){Lovchinsky}, {Slane}, {Gaensler},
  {Hughes}, {Ng}, {Lazendic}, {Gelfand}, \& {Brogan}}]{2011ApJ...731...70L}
{Lovchinsky}, I., {Slane}, P., {Gaensler}, B.~M., {et~al.} 2011, \apj, 731, 70

\bibitem[{{Maeda} {et~al.}(2009){Maeda}, {Uchiyama}, {Bamba}, {Kosugi},
  {Tsunemi}, {Helder}, {Vink}, {Kodaka}, {Terada}, {Fukazawa}, {Hiraga},
  {Hughes}, {Kokubun}, {Kouzu}, {Matsumoto}, {Miyata}, {Nakamura}, {Okada},
  {Someya}, {Tamagawa}, {Tamura}, {Totsuka}, {Tsuboi}, {Ezoe}, {Holt},
  {Ishida}, {Kamae}, {Petre}, \& {Takahashi}}]{2009PASJ...61.1217M}
{Maeda}, Y., {Uchiyama}, Y., {Bamba}, A., {et~al.} 2009, \pasj, 61, 1217

\bibitem[{{Misanovic} {et~al.}(2010){Misanovic}, {Kargaltsev}, \&
  {Pavlov}}]{2010ApJ...725..931M}
{Misanovic}, Z., {Kargaltsev}, O., \& {Pavlov}, G.~G. 2010, \apj, 725, 931

\bibitem[{{Monet} {et~al.}(2003){Monet}, {Levine}, {Canzian}, {Ables}, {Bird},
  {Dahn}, {Guetter}, {Harris}, {Henden}, {Leggett}, {Levison}, {Luginbuhl},
  {Martini}, {Monet}, {Munn}, {Pier}, {Rhodes}, {Riepe}, {Sell}, {Stone},
  {Vrba}, {Walker}, {Westerhout}, {Brucato}, {Reid}, {Schoening}, {Hartley},
  {Read}, \& {Tritton}}]{2003AJ....125..984M}
{Monet}, D.~G., {Levine}, S.~E., {Canzian}, B., {et~al.} 2003, \aj, 125, 984

\bibitem[{{Nobukawa} {et~al.}(2016){Nobukawa}, {Uchiyama}, {Nobukawa},
  {Yamauchi}, \& {Koyama}}]{2016ApJ...833..268N}
{Nobukawa}, M., {Uchiyama}, H., {Nobukawa}, K.~K., {Yamauchi}, S., \& {Koyama},
  K. 2016, \apj, 833, 268

\bibitem[{{Olive} \& {Particle Data Group}(2014)}]{2014ChPhC..38i0001O}
{Olive}, K.~A., \& {Particle Data Group}. 2014, Chinese Physics C, 38, 090001

\bibitem[{{Ozawa} {et~al.}(2009){Ozawa}, {Koyama}, {Yamaguchi}, {Masai}, \&
  {Tamagawa}}]{2009ApJ...706L..71O}
{Ozawa}, M., {Koyama}, K., {Yamaguchi}, H., {Masai}, K., \& {Tamagawa}, T.
  2009, \apjl, 706, L71

\bibitem[{{Pannuti} {et~al.}(2014){Pannuti}, {Rho}, {Heinke}, \&
  {Moffitt}}]{2014AJ....147...55P}
{Pannuti}, T.~G., {Rho}, J., {Heinke}, C.~O., \& {Moffitt}, W.~P. 2014, \aj,
  147, 55

\bibitem[{{Parizot} {et~al.}(2006){Parizot}, {Marcowith}, {Ballet}, \&
  {Gallant}}]{2006A&A...453..387P}
{Parizot}, E., {Marcowith}, A., {Ballet}, J., \& {Gallant}, Y.~A. 2006, \aap,
  453, 387

\bibitem[{{Pavlov} \& {Luna}(2009)}]{2009ApJ...703..910P}
{Pavlov}, G.~G., \& {Luna}, G.~J.~M. 2009, \apj, 703, 910

\bibitem[{{Pavlov} {et~al.}(2002){Pavlov}, {Zavlin}, \&
  {Sanwal}}]{2002nsps.conf..273P}
{Pavlov}, G.~G., {Zavlin}, V.~E., \& {Sanwal}, D. 2002, in Neutron Stars,
  Pulsars, and Supernova Remnants, ed. W.~{Becker}, H.~{Lesch}, \&
  J.~{Tr{\"u}mper}, 273

\bibitem[{{Petruk}(2001)}]{2001A&A...371..267P}
{Petruk}, O. 2001, \aap, 371, 267

\bibitem[{{Reach} {et~al.}(2006){Reach}, {Rho}, {Tappe}, {Pannuti}, {Brogan},
  {Churchwell}, {Meade}, {Babler}, {Indebetouw}, \&
  {Whitney}}]{2006AJ....131.1479R}
{Reach}, W.~T., {Rho}, J., {Tappe}, A., {et~al.} 2006, \aj, 131, 1479

\bibitem[{{Revnivtsev} {et~al.}(2009){Revnivtsev}, {Sazonov}, {Churazov},
  {Forman}, {Vikhlinin}, \& {Sunyaev}}]{2009Natur.458.1142R}
{Revnivtsev}, M., {Sazonov}, S., {Churazov}, E., {et~al.} 2009, \nat, 458, 1142

\bibitem[{{Reynolds} \& {Keohane}(1999)}]{1999ApJ...525..368R}
{Reynolds}, S.~P., \& {Keohane}, J.~W. 1999, \apj, 525, 368

\bibitem[{{Rho} \& {Petre}(1998)}]{1998ApJ...503L.167R}
{Rho}, J., \& {Petre}, R. 1998, \apjl, 503, L167

\bibitem[{{Rosen} {et~al.}(2016){Rosen}, {Webb}, {Watson}, {Ballet}, {Barret},
  {Braito}, {Carrera}, {Ceballos}, {Coriat}, {Della Ceca}, {Denkinson},
  {Esquej}, {Farrell}, {Freyberg}, {Gris{\'e}}, {Guillout}, {Heil},
  {Koliopanos}, {Law-Green}, {Lamer}, {Lin}, {Martino}, {Michel}, {Motch},
  {Nebot Gomez-Moran}, {Page}, {Page}, {Page}, {Pakull}, {Pye}, {Read},
  {Rodriguez}, {Sakano}, {Saxton}, {Schwope}, {Scott}, {Sturm}, {Traulsen},
  {Yershov}, \& {Zolotukhin}}]{2015arXiv150407051R}
{Rosen}, S.~R., {Webb}, N.~A., {Watson}, M.~G., {et~al.} 2016, \aap, 590, A1

\bibitem[{{Ryu} {et~al.}(2009){Ryu}, {Koyama}, {Nobukawa}, {Fukuoka}, \&
  {Tsuru}}]{2009PASJ...61..751R}
{Ryu}, S.~G., {Koyama}, K., {Nobukawa}, M., {Fukuoka}, R., \& {Tsuru}, T.~G.
  2009, \pasj, 61, 751

\bibitem[{{Safi-Harb} {et~al.}(2000){Safi-Harb}, {Petre}, {Arnaud}, {Keohane},
  {Borkowski}, {Dyer}, {Reynolds}, \& {Hughes}}]{2000ApJ...545..922S}
{Safi-Harb}, S., {Petre}, R., {Arnaud}, K.~A., {et~al.} 2000, \apj, 545, 922

\bibitem[{{Sedov}(1959)}]{1959sdmm.book.....S}
{Sedov}, L.~I. 1959, {Similarity and Dimensional Methods in Mechanics}

\bibitem[{{Sezer} {et~al.}(2011){Sezer}, {G{\"o}k}, {Hudaverdi}, {Kimura}, \&
  {Ercan}}]{2011MNRAS.415..301S}
{Sezer}, A., {G{\"o}k}, F., {Hudaverdi}, M., {Kimura}, M., \& {Ercan}, E.~N.
  2011, \mnras, 415, 301

\bibitem[{{Skrutskie} {et~al.}(2006){Skrutskie}, {Cutri}, {Stiening},
  {Weinberg}, {Schneider}, {Carpenter}, {Beichman}, {Capps}, {Chester},
  {Elias}, {Huchra}, {Liebert}, {Lonsdale}, {Monet}, {Price}, {Seitzer},
  {Jarrett}, {Kirkpatrick}, {Gizis}, {Howard}, {Evans}, {Fowler}, {Fullmer},
  {Hurt}, {Light}, {Kopan}, {Marsh}, {McCallon}, {Tam}, {Van Dyk}, \&
  {Wheelock}}]{2006AJ....131.1163S}
{Skrutskie}, M.~F., {Cutri}, R.~M., {Stiening}, R., {et~al.} 2006, \aj, 131,
  1163

\bibitem[{{Slane} {et~al.}(2015){Slane}, {Bykov}, {Ellison}, {Dubner}, \&
  {Castro}}]{2015SSRv..188..187S}
{Slane}, P., {Bykov}, A., {Ellison}, D.~C., {Dubner}, G., \& {Castro}, D. 2015,
  \ssr, 188, 187

\bibitem[{{Slane} {et~al.}(1999){Slane}, {Gaensler}, {Dame}, {Hughes},
  {Plucinsky}, \& {Green}}]{1999ApJ...525..357S}
{Slane}, P., {Gaensler}, B.~M., {Dame}, T.~M., {et~al.} 1999, \apj, 525, 357

\bibitem[{{Slane} {et~al.}(2001){Slane}, {Hughes}, {Edgar}, {Plucinsky},
  {Miyata}, {Tsunemi}, \& {Aschenbach}}]{2001ApJ...548..814S}
{Slane}, P., {Hughes}, J.~P., {Edgar}, R.~J., {et~al.} 2001, \apj, 548, 814

\bibitem[{{Slane} {et~al.}(2002){Slane}, {Smith}, {Hughes}, \&
  {Petre}}]{2002ApJ...564..284S}
{Slane}, P., {Smith}, R.~K., {Hughes}, J.~P., \& {Petre}, R. 2002, \apj, 564,
  284

\bibitem[{{Smith} {et~al.}(2001){Smith}, {Brickhouse}, {Liedahl}, \&
  {Raymond}}]{2001ApJ...556L..91S}
{Smith}, R.~K., {Brickhouse}, N.~S., {Liedahl}, D.~A., \& {Raymond}, J.~C.
  2001, \apjl, 556, L91

\bibitem[{{Smith} \& {Hughes}(2010)}]{2010ApJ...718..583S}
{Smith}, R.~K., \& {Hughes}, J.~P. 2010, \apj, 718, 583

\bibitem[{{Spitzer}(1962)}]{1962pfig.book.....S}
{Spitzer}, L. 1962, {Physics of Fully Ionized Gases} (New York: Interscience)

\bibitem[{{Taylor}(1950)}]{1950RSPSA.201..159T}
{Taylor}, G. 1950, Proceedings of the Royal Society of London Series A, 201,
  159

\bibitem[{{Truelove} \& {McKee}(1999)}]{1999ApJS..120..299T}
{Truelove}, J.~K., \& {McKee}, C.~F. 1999, \apjs, 120, 299

\bibitem[{{Truelove} \& {McKee}(2000)}]{2000ApJS..128..403T}
---. 2000, \apjs, 128, 403

\bibitem[{{Tsubone} {et~al.}(2016){Tsubone}, {Sawada}, {Bamba}, {Katsuda}, \&
  {Vink}}]{2016arXiv161201221T}
{Tsubone}, Y., {Sawada}, M., {Bamba}, A., {Katsuda}, S., \& {Vink}, J. 2016,
  ApJ, in press, arXiv:1612.01221

\bibitem[{{Uchida} {et~al.}(2012){Uchida}, {Tsunemi}, {Katsuda}, {Mori},
  {Petre}, \& {Yamaguchi}}]{2012PASJ...64...61U}
{Uchida}, H., {Tsunemi}, H., {Katsuda}, S., {et~al.} 2012, \pasj, 64

\bibitem[{{Uchida} {et~al.}(2013){Uchida}, {Yamaguchi}, \&
  {Koyama}}]{2013ApJ...771...56U}
{Uchida}, H., {Yamaguchi}, H., \& {Koyama}, K. 2013, \apj, 771, 56

\bibitem[{{Uchiyama} {et~al.}(2013){Uchiyama}, {Nobukawa}, {Tsuru}, \&
  {Koyama}}]{2013PASJ...65...19U}
{Uchiyama}, H., {Nobukawa}, M., {Tsuru}, T.~G., \& {Koyama}, K. 2013, \pasj,
  65, 19

\bibitem[{{Vink}(2008)}]{2008A&A...486..837V}
{Vink}, J. 2008, \aap, 486, 837

\bibitem[{{Vink}(2012)}]{2012A&ARv..20...49V}
---. 2012, \aapr, 20, 49

\bibitem[{{Vink} \& {Laming}(2003)}]{2003ApJ...584..758V}
{Vink}, J., \& {Laming}, J.~M. 2003, \apj, 584, 758

\bibitem[{{Voges} {et~al.}(1999){Voges}, {Aschenbach}, {Boller},
  {Br{\"a}uninger}, {Briel}, {Burkert}, {Dennerl}, {Englhauser}, {Gruber},
  {Haberl}, {Hartner}, {Hasinger}, {K{\"u}rster}, {Pfeffermann}, {Pietsch},
  {Predehl}, {Rosso}, {Schmitt}, {Tr{\"u}mper}, \&
  {Zimmermann}}]{1999A&A...349..389V}
{Voges}, W., {Aschenbach}, B., {Boller}, T., {et~al.} 1999, \aap, 349, 389

\bibitem[{{Warren} {et~al.}(2005){Warren}, {Hughes}, {Badenes}, {Ghavamian},
  {McKee}, {Moffett}, {Plucinsky}, {Rakowski}, {Reynoso}, \&
  {Slane}}]{2005ApJ...634..376W}
{Warren}, J.~S., {Hughes}, J.~P., {Badenes}, C., {et~al.} 2005, \apj, 634, 376

\bibitem[{{White} \& {Long}(1991)}]{1991ApJ...373..543W}
{White}, R.~L., \& {Long}, K.~S. 1991, \apj, 373, 543

\bibitem[{{Whiteoak} \& {Green}(1996)}]{1996A&AS..118..329W}
{Whiteoak}, J.~B.~Z., \& {Green}, A.~J. 1996, \aaps, 118, 329

\bibitem[{{Wilms} {et~al.}(2000){Wilms}, {Allen}, \&
  {McCray}}]{2000ApJ...542..914W}
{Wilms}, J., {Allen}, A., \& {McCray}, R. 2000, \apj, 542, 914

\bibitem[{{Yamaguchi} {et~al.}(2009){Yamaguchi}, {Ozawa}, {Koyama}, {Masai},
  {Hiraga}, {Ozaki}, \& {Yonetoku}}]{2009ApJ...705L...6Y}
{Yamaguchi}, H., {Ozawa}, M., {Koyama}, K., {et~al.} 2009, \apjl, 705, L6

\bibitem[{{Yamauchi} {et~al.}(2013){Yamauchi}, {Nobukawa}, {Koyama}, \&
  {Yonemori}}]{2013PASJ...65....6Y}
{Yamauchi}, S., {Nobukawa}, M., {Koyama}, K., \& {Yonemori}, M. 2013, \pasj,
  65, 6

\bibitem[{{Yamauchi} {et~al.}(2014){Yamauchi}, {Shimizu}, {Nakashima},
  {Nobukawa}, {Tsuru}, \& {Koyama}}]{2014PASJ...66..125Y}
{Yamauchi}, S., {Shimizu}, M., {Nakashima}, S., {et~al.} 2014, \pasj, 66, 125

\bibitem[{{Yamauchi} {et~al.}(2008){Yamauchi}, {Ueno}, {Koyama}, \&
  {Bamba}}]{2008PASJ...60.1143Y}
{Yamauchi}, S., {Ueno}, M., {Koyama}, K., \& {Bamba}, A. 2008, \pasj, 60, 1143

\bibitem[{{Yuasa} {et~al.}(2012){Yuasa}, {Makishima}, \&
  {Nakazawa}}]{2012ApJ...753..129Y}
{Yuasa}, T., {Makishima}, K., \& {Nakazawa}, K. 2012, \apj, 753, 129

\bibitem[{{Zhou} {et~al.}(2014){Zhou}, {Safi-Harb}, {Chen}, {Zhang}, {Jiang},
  \& {Ferrand}}]{2014ApJ...791...87Z}
{Zhou}, P., {Safi-Harb}, S., {Chen}, Y., {et~al.} 2014, \apj, 791, 87

\end{thebibliography}
